\def\paperformat{ieee}  % Change to 'ieee' for IEEE standard format
\ifx\paperformat\undefined
\def\paperformat{tmlcn}  % Default to tmlcn if not set
\fi

\def\ieeeformat{ieee}
\def\tmlcnformat{tmlcn}

\ifx\paperformat\ieeeformat
\documentclass[journal]{IEEEtran} %,onecolumn,12pt

\else
\documentclass{ieeetmlcn}
\fi

\usepackage{color}
\usepackage{hyperref}
\usepackage{setspace}
\usepackage{times}
\usepackage[pdftex]{graphicx}
\usepackage{latexsym}
\usepackage{dsfont}
\usepackage{cite}
\usepackage{verbatim}
\usepackage{subfigure}
\usepackage[acronym,nogroupskip,nonumberlist,nopostdot]{glossaries}
\usepackage{xspace}
\usepackage{cite}
\usepackage{amsmath,amssymb,amsfonts}
\usepackage{algorithmic}
\usepackage{algorithm}
\usepackage{graphicx}
\usepackage{textcomp}
\usepackage{xcolor}
\usepackage{float}

\newcommand{\secref}[1]{{Sec.}~\ref{#1}}
\newcommand{\figref}[1]{{Fig.}~\ref{#1}}

\def\bb0{{\mathbb{0}}}

\def\ba{{\mathbf{a}}}
\def\bb{{\mathbf{b}}}

\def\bn{{\mathbf{n}}}

\def\bw{{\mathbf{w}}}
\def\bx{{\mathbf{x}}}
\def\by{{\mathbf{y}}}
\def\bz{{\mathbf{z}}}
\def\b0{{\mathbf{0}}}
\def\bA{{\mathbf{A}}}
\def\bB{{\mathbf{B}}}

\def\bG{{\mathbf{G}}}
\def\bH{{\mathbf{H}}}

\def\bT{{\mathbf{T}}}

\def\bX{{\mathbf{X}}}

\def\cA{\mathcal{A}}

\def\cD{\mathcal{D}}

\def\cF{\mathcal{F}}
\def\cG{\mathcal{G}}

\def\cM{\mathcal{M}}

\def\cR{\mathcal{R}}
\def\cS{\mathcal{S}}

\def\cW{\mathcal{W}}

\def\sf0{{\mathsf{0}}}
\def\rm0{{\mathrm{0}}}
\def\b0{{\pmb{0}}} 
\def\vec{\mathrm{vec}~}
\def\kron{\otimes}

\newcommand{\trace}[1]{\textrm{Tr}({#1})}

\newacronym{wlan}{WLAN}{wireless local area network}

\newacronym{manet}{MANET}{mobile ad hoc network}

\newacronym{dma}{DMA}{dynamic metasurface antenna}

\newacronym{dmas}{DMAs}{dynamic metasurface antennas}

\newacronym{upa}{UPA}{uniform planar array}

\newacronym{rf}{RF}{radio frequency}

\newacronym{mimo}{MIMO}{multiple input-multiple output}

\newacronym{mmwave}{mmWave}{millimeter wave}

\newacronym{mmwavemimo}{mmWave MIMO}{millimeter wave multiple input-multiple output}

\newacronym{ue}{UE}{user equipment}

\newacronym{ues}{UEs}{user equipments}

\newacronym{bs}{BS}{base station}

\newacronym{em}{EM}{electromagnetic}

\newacronym{ofdm}{OFDM}{orthogonal frequency-division multiplexing}

\newacronym{pas}{PAS}{power angular spectrum}

\newacronym{gcs}{GCS}{global coordinate system}

\newacronym{lcs}{LCS}{local coordinate system}

\newcommand{\figsize}{0.85}

\newcommand{\Gs}{G_{\text{s}}} % spatial grid size
\newcommand{\Mgrid}{H} % number of hierarchical levels
\newcommand{\gridindex}{h}
\newcommand{\gridenc}{\mathcal{E}} % grid encoding function
\newcommand{\gridtensor}{\mathcal{O}} % hierarchical grid tensor
\newcommand{\gridlevel}[1]{\mathcal{\gridtensor}^{(#1)}} % grid at level m
\newcommand{\dgrid}{d_{\text{grid}}} % grid feature dimension
\newcommand{\hlevel}[1]{\mathbf{z}^{(#1)}} % per-level features
\newcommand{\hloc}{\mathbf{z}_{\text{loc}}} % combined location features
\newcommand{\Lcount}{\mathcal{L}_1} % Stage 1 loss
\newcommand{\Lkl}{\mathcal{L}_{\text{KL}}} % KL divergence loss
\newcommand{\Ltotal}{\mathcal{L}_{\text{count}}} % total count loss
\newcommand{\Lsparse}{\mathcal{L}_{\text{sparse}}} % sparsity loss
\newcommand{\lambdakl}{\lambda_{\text{KL}}} % KL weight
\newcommand{\lambdatotal}{\lambda_{\text{count}}} % total count weight
\newcommand{\lambdasparse}{\lambda_{\text{sparse}}} % sparsity weight
\newcommand{\taus}{\rho_s} % sparsity threshold
\newcommand{\dcontext}{d_{\text{context}}} % context dimension
\newcommand{\Ltwo}{\mathcal{L}_2} % Stage 2 loss
\newcommand{\Lthree}{\mathcal{L}_3} % Stage 3 loss
\newcommand{\Lmag}{\mathcal{L}_{\text{mag}}} % magnitude loss
\newcommand{\Ldir}{\mathcal{L}_{\text{dir}}} % direction loss
\newcommand{\lambdamag}{\lambda_{\text{mag}}} % magnitude weight
\definecolor{mygreen}{rgb}{0 0.5 0}
\newcommand{\revfive}[1]{{\color{black}{#1}}}

\newcommand{\Tcoh}{T_\text{coh}}

\newcommand{\Hk}{\mathbf{H}[k]}

\newcommand{\Nry}{N_\text{\rx,y}}
\newcommand{\Nrz}{N_\text{\rx,z}}

\newcommand{\azi}{\phi}
\newcommand{\ele}{\theta}
\newcommand{\tx}{\text{t}}
\newcommand{\rx}{\text{r}}
\newcommand{\col}{\text{C}}
\newcommand{\cross}{\text{X}}

\newcommand{\polAng}{\rho}
\newcommand{\orientVec}{\Theta}

\newcommand{\trans}[1]{#1^{\mathsf{T}}}
\newcommand{\herm}[1]{#1^{*}} % conjugate transpose (Hermitian)
\newcommand{\conj}[1]{\overline{#1}} % conjugate (bar notation)
\newcommand{\rot}[1]{\text{R}_{\text{#1}}}
\newcommand{\rotPol}{\rot{p}(\polAng)}

\newcommand{\rotComb}{\rot{}}

\let\vec\relax      % Undefine \vec
\DeclareMathAccent{\vec}{\mathord}{letters}{"7E}  % Restore arrow accent

\newcommand{\unitX}{\vec{\bx}}
\newcommand{\unitY}{\vec{\by}}
\newcommand{\unitZ}{\vec{\bz}}

\newcommand{\unitr}{\vec{\mathbf{r}}}
\newcommand{\unitPhi}{\vec{\boldsymbol{\phi}}}
\newcommand{\unitTheta}{\vec{\boldsymbol{\theta}}}

\newcommand{\Real}{\mathbb{R}}
\newcommand{\Complex}{\mathbb{C}}

\newcommand{\GtoLsymb}{Q}
\newcommand{\GtoL}{\GtoLsymb(\theta,\phi,\orientVec)}

\newcommand{\Npath}{L}
\newcommand{\Npathall}{\Npath_\text{all}}
\newcommand{\pgain}{\alpha_\ell}
\newcommand{\dod}{\theta_{\ell,\text{t}},\phi_{\ell,\text{t}}}
\newcommand{\doa}{\theta_{\ell,\text{r}},\phi_{\ell,\text{r}}}
\newcommand{\pdepol}{\bX}
\newcommand{\pdelay}{\tau_\ell}
\newcommand{\pulseshape}{g}

\newcommand{\onebytwo}[2]{\begin{bmatrix}
		#1 & #2 
\end{bmatrix} }

\newcommand{\Nmeas}{M}
\newcommand{\angleset}{\cA}
\newcommand{\measset}{\cM}
\newcommand{\matrixset}{\cR}
\newcommand{\optfuncname}{f}

\newcommand{\rsrp}{\text{RSRP}}
\newcommand{\powsig}{P_{\text{s}}}
\newcommand{\wpanel}{\mathbf{w}_p}
\newcommand{\Hpanel}{\mathbf{H}_p}
\newcommand{\bft}{\mathbf{f}} % transmit beamformer
\newcommand{\pathgain}{\beta_{\ell,k}}
\newcommand{\Grx}{\mathbf{G}_{\rx,\ell}}
\newcommand{\Gtx}{\mathbf{G}_{\tx,\ell}}
\newcommand{\brx}{\mathbf{g}_{\rx,\ell}}
\newcommand{\btx}{\mathbf{g}_{\tx,\ell}}
\newcommand{\amat}{\mathbf{a}}

\newcommand{\bvec}{\mathbf{b}}
\newcommand{\xvec}{\mathbf{x}}
\newcommand{\Rmatrix}{\mathbf{R}}
\newcommand{\Nbeam}{N_\text{b}}

\newcommand{\Ngrid}{G}
\newcommand{\Nusers}{N}
\newcommand{\xloc}{\mathbf{v}}

\newcommand{\cvec}{\mathbf{c}}
\newcommand{\pvec}{\mathbf{p}}

\newcommand{\Lvec}{\mathbf{L}}

\newcommand{\onehot}[1]{\mathbf{e}_{#1}}

\newcommand{\ctx}{\mathbf{c}_{\text{context}}}
\newcommand{\uvec}{\mathbf{u}}

\newcommand{\pathinfo}{path information}

\DeclareMathOperator*{\argmin}{arg\,min}

\ifx\paperformat\tmlcnformat
\usepackage{algorithm}
\fi

\ifx\paperformat\ieeeformat
\fi

\ifx\paperformat\tmlcnformat
\def\BibTeX{{\rm B\kern-.05em{\sc i\kern-.025em b}\kern-.08em
		T\kern-.1667em\lower.7ex\hbox{E}\kern-.125emX}}
\AtBeginDocument{\definecolor{tmlcncolor}{cmyk}{0.93,0.59,0.15,0.02}\definecolor{NavyBlue}{RGB}{0,86,125}}

\def\OJlogo{\vspace{-4pt}\includegraphics[height=18pt]{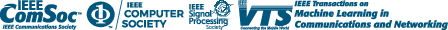}}
\def\seclogo{\vspace{10pt}\includegraphics[height=18pt]{new_logo.eps}}

\else

\IEEEoverridecommandlockouts
\usepackage[top=0.71in,left=0.63in,right=0.63in,bottom=1.04in]{geometry}

\def\BibTeX{{\rm B\kern-.05em{\sc i\kern-.025em b}\kern-.08em
		T\kern-.1667em\lower.7ex\hbox{E}\kern-.125emX}}
\fi

\begin{document}
	
	% ============================================================================
	% TMLCN-SPECIFIC METADATA
	% ============================================================================
	\ifx\paperformat\tmlcnformat
%	\receiveddate{XX Month, XXXX}
 	\reviseddate{11 February, 2026}
	% \accepteddate{XX Month, XXXX}
	% \publisheddate{XX Month, XXXX}
	% \currentdate{XX Month, XXXX}
	% \doiinfo{TMLCN.2022.1234567}
	% \markboth{}{Author {et al.}}
	\fi
	
	% ============================================================================
	% TITLE
	% ============================================================================
	\title{%
		\vspace{-0.25em}%
		{\footnotesize\normalfont\raggedright This work has been submitted to the IEEE for possible publication. 
			Copyright may be transferred without notice, after which this version may no longer be accessible.\par}%
		\vspace{0.5em}%
		Heterogeneity-agnostic AI/ML-assisted beam selection for multi-panel arrays%
	}
	
	% ============================================================================
	% AUTHORS - Format depends on paper format
	% ============================================================================
	\ifx\paperformat\ieeeformat
	% IEEE Standard Author Block
	\author{\IEEEauthorblockN{Ibrahim Kilinc \textit{Graduate Student Member, IEEE} and Robert W. Heath Jr., \textit{Fellow, IEEE}}

	\thanks{Ibrahim Kilinc  and Robert W.  Heath Jr. are with the 
			Department of Electrical and Computer Engineering, University of California San Diego, La Jolla, CA 92093 USA (e-mail: ikilinc@ucsd.edu;  rwheathjr@ucsd.edu).
			 This material is based upon work supported in part by the National Science Foundation under grant no NSF-CNS-2433782 and is supported in part by funds from federal agency and industry partners as specified in the Resilient \& Intelligent NextG Systems (RINGS) program.}
	}

%		
%	\IEEEauthorblockA{ECE Department, University of California, San Diego, USA}\\
%		% (e-mail: \{ikilinc, rwheathjr\}@ucsd.edu)}
%	}
%	

	\maketitle

	\else
	% TMLCN Author Block
%	\author{%
%		First Author\authorrefmark{1},
%		Second Author\authorrefmark{2}
%		\thanks{\authorrefmark{1}Department, University, City, Country (e-mail: first.author@email.com)}
%		\thanks{\authorrefmark{2}Department, University, City, Country (e-mail: second.author@email.com)}
%	}
	\author{IBRAHIM KILINC (Graduate Student Member, IEEE), \\ AND ROBERT W. HEATH JR. (Fellow, IEEE)}
	\affil{Department of Electrical and Computer Engineering, University of California, San Diego, USA}
	\authornote{``This material is based upon work supported in part by the National Science Foundation under grant no NSF-CNS-2433782 and is supported in part by funds from federal agency and industry partners as specified in the Resilient \& Intelligent NextG Systems (RINGS) program.''}
	\fi
	
	% ============================================================================
	% ABSTRACT
	% ============================================================================
	\begin{abstract}
		
		%				Configuring beamforming weights in MIMO communication is crucial to achieve higher data rates for 5G and beyond. 
		AI/ML-based beam selection methods coupled with location information effectively reduce beam training overhead. Unfortunately, heterogeneous antenna hardware with varying dimensions, orientations, codebooks, element patterns, and polarization angles limits their feasibility and generalization. This challenge requires either a heterogeneity-agnostic model functional under these variations, or developing many models for each configuration, which is infeasible and expensive in practice. In this paper, we propose a unifying AI/ML-based beam selection algorithm supporting antenna heterogeneity by predicting wireless propagation characteristics independent of antenna configuration. We derive a reference signal received power (RSRP) model that decouples propagation characteristics from antenna configuration. We propose an optimization framework to extract propagation variables consisting of angle-of-arrival (AoA), angle-of-departure (AoD), and a matrix incorporating path gain and channel depolarization from beamformed RSRP measurements. We develop a three-stage autoregressive network to predict these variables from user location, enabling RSRP calculation and beam selection for arbitrary antenna configurations without retraining or having a separate model for each configuration. Simulation results show our heterogeneity-agnostic method provides spectral efficiency close to that of genie-aided selection both with and without antenna heterogeneity.
	\end{abstract}

	% ============================================================================
	% KEYWORDS
	% ============================================================================
	\begin{IEEEkeywords}
	% Enter key words or phrases in alphabetical order, separated by
		% commas. Using the \textit{IEEE Thesaurus} can help you find the best
		% standardized keywords to fit your article. Use the \underline{\href{https://www.ieee.org/publications/services/thesaurus.html}{thesaurus
				% access request form}} for free access to the \textit{IEEE Thesaurus}.
	Antenna heterogeneity, multi-panel arrays, beam selection, AI/ML-assisted
	\end{IEEEkeywords}
	
	% ============================================================================
	% SPECIAL PAPER NOTICE (Optional, TMLCN only)
	% ============================================================================
	% \IEEEspecialpapernotice{(Invited Paper)}
	
	\maketitle
	
	\section{INTRODUCTION}\label{sec: introduction}

	\IEEEPARstart{M}{IMO} is a crucial technology to support higher peak data rates, enhanced spectrum efficiency and better coverage for sixth generation cellular networks (6G), and beyond. Configuring MIMO links, however, is challenging with the increasing MIMO dimensions \cite{HengEtAlSixKeyChallengesBeam2021,Gonzalez-PrelcicEtAlIntegratedSensingCommunicationRevolution2024}. Multi-panel antenna arrays with increasing number of elements introduce finer beam resolution with increasing gains. The receive and transmit array configuration, \revfive{however}, causes greater overhead in beam-based approaches \cite{Gonzalez-PrelcicEtAlIntegratedSensingCommunicationRevolution2024,HengEtAlSixKeyChallengesBeam2021,QurratulainKhanEtAlMachineLearningMillimeterWave2023}. Furthermore, user mobility requires quicker array configuration due to shorter channel coherence duration \cite{HengEtAlSixKeyChallengesBeam2021,IqbalEtAlMobilityAnalysisUESideBeamforming2023,QurratulainKhanEtAlMachineLearningMillimeterWave2023}. Therefore, devices equipped with multi-panel arrays in mobile environment require scalable, lightweight and fast beamforming solutions to handle complex MIMO configurations.

	Beam training over a set of beam weights, known as a \revfive{beam} codebook, is well \revfive{used} in the standards and also in the research community. Larger arrays with finer beams require larger codebooks to cover the angular space and in turn brute force search over whole codebook becomes infeasible \cite{HengEtAlSixKeyChallengesBeam2021}. In 5G, beam management involves several stages with coarse beam search, feedback and beam refinement \cite{3GPP_TR38901}. It is a multi-stage beam training that searches over predefined codebooks \cite{DreifuerstHeathMassiveMIMO5GHow2023,3GPP_TR38901}. Prior work on beam training focuses on hierarchical search as in 5G, AI/ML-based beam selection leveraging side information to decrease beam training overhead \cite{Gonzalez-PrelcicEtAlIntegratedSensingCommunicationRevolution2024,QurratulainKhanEtAlMachineLearningMillimeterWave2023,RezaieEtAlDeepLearningApproachLocation2022,LiEtAlDeepReinforcementLearningBasedMultiPanel2021,kilincasilomardma,Reus-MunsEtAlDeepLearningVisualLocation2021}. Almost all of these methods assume fixed antenna configurations: a single panel with fixed size, fixed polarization angle, specific codebooks, defined antenna patterns, and fixed orientations \cite{LiEtAlDeepReinforcementLearningBasedMultiPanel2021,WangEtAlMmWaveVehicularBeamSelection2019,Reus-MunsEtAlDeepLearningVisualLocation2021}. User equipments (UEs), base stations (BSs) in practice, however, \revfive{are heterogeneous in that it supports a diverse set of antenna configurations \cite{CuiEtAlScalableVideoMulticastMUMIMO2016}. As a result, the approaches in prior work that assume fixed antenna configurations do not necessarily generalize when the hardware changes}.

	Heterogeneity in the context of beam selection refers to the variations in antenna hardware and deployment setups across devices. These variations include the number of antenna panels, their placement on the device, panel sizes, orientations, element patterns, polarization types and beam codebooks. Such variations are common in wireless systems. \revfive{For example, handsets, and vehicles} all have different antenna layouts \cite{HengEtAlSixKeyChallengesBeam2021,CuiEtAlScalableVideoMulticastMUMIMO2016,LiEtAlDeepReinforcementLearningBasedMultiPanel2021}. 5G beam management is flexible to operate in these variations \cite{DreifuerstHeathMassiveMIMO5GHow2023} since there is no specifically pretrained \revfive{algorithm} in the protocol. AI/ML-based beam selection solutions in prior work, however, depend on these variations due to fixed input-output dimensions, being tailored to a fixed codebook, orientation and antenna size \cite{goodfellow2016deep,PatelHeathHarnessingMultimodalSensingMultiuser2024,WangEtAlMmWaveVehicularBeamSelection2019,Reus-MunsEtAlDeepLearningVisualLocation2021}. One approach to solving this problem is to train separate models for each possible antenna configuration. Unfortunately, this approach becomes infeasible as device diversity increases and new devices are released, since it would require maintaining a separate model for each manufacturer's device. Therefore, AI/ML-based solutions in prior work may not be directly applicable to handle such heterogeneity.

	\revfive{In this work, we propose an AI/ML-assisted, location-based, heterogeneity-agnostic multi-panel beam selection method that overcomes the multi-panel beam configuration and antenna heterogeneity problem}. We consider a BS equipped with a uniform planar array (UPA) with phase shifters. The UEs have multi-panel UPAs, where each panel points in a different direction. We assume a single panel is selected at a time through panel switching. Unlike conventional methods, our method relies on a heterogeneity-agnostic wireless propagation information for the AI/ML model, that we name as \pathinfo. The angular {\pathinfo} is a set, where each element belongs to a ray with AoA/AoD and combined path matrix characterizing channel depolarization, and path gains. Path information is independent of antenna heterogeneity, which is the foundation of our heterogeneity-agnostic beam selection model. We derive an RSRP expression decoupling antenna configuration and propagation characteristics. We then formulate an optimization problem to solve the path information from RSRP measurements for a given antenna configuration. We develop a three-stage model consisting of convolutional feature extraction, autoregressive AoA/AoD prediction and dense layers, which maps path information with UE locations. We provide extensive simulation results that highlight the significance of heterogeneity consideration as well as performance gains. Our contributions can be summarized as follows:
	
	\begin{itemize}
		\item We derive an analytical RSRP model that parametrizes antenna heterogeneity and propagation into separate variables. The model decouples heterogeneity aspects and propagation information, enabling heterogeneity-agnostic beam and panel selection with given path variables. It eliminates the need of retraining/developing predictors for every possible antenna configuration.
		
		\item We formulate an optimization problem to extract path information from RSRP measurements with known heterogeneity configurations and user locations. We propose a two-stage alternating optimization: the first stage for gradient-free AoA and AoD optimization, and the second stage for combined path matrix optimization through semidefinite programming. The extracted path information incorporates angular gain and channel depolarization, and is agnostic to antenna heterogeneity.
		\item We develop a novel three-stage neural network that predicts path information from UE locations. The network addresses the permutation invariance of paths through a count-based decomposition with autoregressive prediction. Our heterogeneity-agnostic method, \textit{Static Pred}, leverages the predicted path information to calculate RSRPs for any antenna configuration.
		
		\item We ray-trace channels in \revfive{a vehicular environment with blockages}. The RSRP measurements are obtained, the path information is optimized. The path predictor network is trained with location and path information.
		
	\end{itemize}

	There is vast prior work on sensor-aided beam selection methods, specifically on location-based approaches \cite{PatelHeathHarnessingMultimodalSensingMultiuser2024,WangEtAlMmWaveVehicularBeamSelection2019,Reus-MunsEtAlDeepLearningVisualLocation2021,AbdElMoatyMohamedGoudaEtAlEnhancedPositionAidedBeamPrediction2025,KhanEtAlDigitalTwinAssistedExplainableAI2025,AliEtAlOrientationAssistedBeamManagement5G2021,NguyenEtAlBeamManagementOrientationRSRP2022}. The prior work on beam and panel configuration for multi-panel arrays, however, \revfive{was} limited \cite{RezaieEtAlDeepLearningApproachLocation2022,LiEtAlDeepReinforcementLearningBasedMultiPanel2021,kilincasilomardma}. We review relevant prior work in two categories: 1) the work related to beam and panel configuration for multi-panel arrays, and 2) the \revfive{work} on single panel beam selection using location information. We start reviewing the work on multi-panel beam configuration \cite{RezaieEtAlDeepLearningApproachLocation2022,LiEtAlDeepReinforcementLearningBasedMultiPanel2021,kilincasilomardma}. A neural network (NN) in \cite{RezaieEtAlDeepLearningApproachLocation2022} \revfive{predicted} the received signal power for all beams across all panels based on location and orientation information in an indoor environment. This approach, however, \revfive{was} limited to indoor scenarios with fixed codebooks, antenna pattern and sizes. A deep RL framework in \cite{LiEtAlDeepReinforcementLearningBasedMultiPanel2021} \revfive{configured} multi-panels at a multi-sector BS, but it \revfive{neglected} the size, orientation of UE antenna, and codebooks besides simulations limited to statistical channels. Our prior work \cite{kilincasilomardma} \revfive{had} a quasi-heterogeneity-agnostic method for multi-panel dynamic metasurface antennas. The approach \revfive{optimized} angular power grids \revfive{not agnostic to antenna polarization, element patterns, and indirectly orientation}. Consequently, existing multi-panel beam configuration work \revfive{had} limitations under antenna heterogeneity in realistic deployments.

	There is rich prior work on location-based AI/ML-aided beam selection for single panel antenna arrays \cite{PatelHeathHarnessingMultimodalSensingMultiuser2024,WangEtAlMmWaveVehicularBeamSelection2019,Reus-MunsEtAlDeepLearningVisualLocation2021,AbdElMoatyMohamedGoudaEtAlEnhancedPositionAidedBeamPrediction2025,KhanEtAlDigitalTwinAssistedExplainableAI2025,AliEtAlOrientationAssistedBeamManagement5G2021,NguyenEtAlBeamManagementOrientationRSRP2022}. The location is useful as \revfive{similar} positions in the environment observes the same static obstacles for propagation. While dynamic reflectors can vary, there is still a subset of paths and therefore beam combinations \revfive{thus} tend to work well as a function of location. These methods significantly reduce beam training overhead since AI/ML methods are well-suited to model implicit relationships between sensor data and wireless characteristics \cite{PatelHeathHarnessingMultimodalSensingMultiuser2024,DreifuerstEtAlContextawareCodebookDesign6G}. Situational-awareness in the form of occupancy grids obtained by sensor information in vehicles \revfive{was} exploited to select beam subsets using decision tree based approach in \cite{WangEtAlMmWaveVehicularBeamSelection2019}. UE location information \revfive{was} used in a fusion network to infer beam pair subsets in \cite{Reus-MunsEtAlDeepLearningVisualLocation2021}. Signal power measurements from wide beams \revfive{were} used to predict narrow beams from an oversampled DFT codebook in \cite{KhanEtAlDigitalTwinAssistedExplainableAI2025}. Beam indices \revfive{were} predicted from GPS coordinates in \cite{AbdElMoatyMohamedGoudaEtAlEnhancedPositionAidedBeamPrediction2025}, where models \revfive{were} trained separately for different configurations. The best UE beam of a single antenna panel \revfive{was} predicted using prior beam power measurements and antenna orientation in \cite{AliEtAlOrientationAssistedBeamManagement5G2021}. The best beam of a UE with a single antenna panel \revfive{was} selected via beam power predictions based on antenna orientation and previous beam measurements in \cite{NguyenEtAlBeamManagementOrientationRSRP2022}, where the method \revfive{was} trained on a large dataset. The work on single panel \cite{WangEtAlMmWaveVehicularBeamSelection2019,Reus-MunsEtAlDeepLearningVisualLocation2021,KhanEtAlDigitalTwinAssistedExplainableAI2025,AbdElMoatyMohamedGoudaEtAlEnhancedPositionAidedBeamPrediction2025,AliEtAlOrientationAssistedBeamManagement5G2021,NguyenEtAlBeamManagementOrientationRSRP2022}, however, mainly \revfive{proposed} approaches tailored to fixed antenna configurations and \revfive{did} not consider generalizations in realistic deployments with severe antenna heterogeneity.

	In this paper, we develop a heterogeneity-agnostic beam selection methodology for user devices equipped with multi-panel antenna arrays. Compared to prior work \cite{RezaieEtAlDeepLearningApproachLocation2022,LiEtAlDeepReinforcementLearningBasedMultiPanel2021,kilincasilomardma,WangEtAlMmWaveVehicularBeamSelection2019,Reus-MunsEtAlDeepLearningVisualLocation2021,KhanEtAlDigitalTwinAssistedExplainableAI2025,AbdElMoatyMohamedGoudaEtAlEnhancedPositionAidedBeamPrediction2025,AliEtAlOrientationAssistedBeamManagement5G2021,NguyenEtAlBeamManagementOrientationRSRP2022}, our method is fully agnostic to antenna heterogeneity including variations in antenna size, element pattern, orientation, polarization angle, and codebooks. Our formulation decouples antenna configurations from propagation characteristics, which do not change with respect to antenna heterogeneity. Our simulation results show that our method achieves spectral efficiency close to genie-aided spectral efficiency under various levels of antenna heterogeneity.

	\textbf{Notation:} $\mathbf{A}$ is a matrix, $\mathbf{a}$ is a column vector, and $a$, $A$ are scalars. $\trans{\mathbf{A}}$, $\conj{\mathbf{A}}$, $\mathbf{A}^*$ represent the transpose, conjugate, and conjugate transpose. $\|\mathbf{a}\|_2 = \mathbf{a}^*\mathbf{a}$ is for vectors. $|\bA|$, $\kron$, $\text{vec}(\bA)$, and $\trace{\bA}$ denote magnitude, Kronecker product, vectorization, and trace operation. $\vec{\mathbf{x}}$ denotes a unit vector for the basis of a coordinate system and $\hat{(\cdot)}$ denotes an estimate.

	The remainder of the paper is organized as follows. We define the system model to introduce coordinate systems, antenna array structure, channel model, and received signal model in \secref{sec: system_model}. In \secref{sec:rsrp_agnostic}, we derive the RSRP model and introduce the optimization framework to solve propagation characteristics from RSRP measurements. Next, we present our three-stage path information predictor and multi-panel beam selection in \secref{sec:het_agnostic_section}. Lastly, we present extensive simulations in \secref{sec:simulation} and conclude the manuscript in \secref{sec:conclusion}.	
	
	\section{SYSTEM MODEL}\label{sec: system_model}
	Our heterogeneity-agnostic beam selection explicitly separates antenna configuration from propagation characteristics. This separation enables independent parameterization of antenna configuration from the underlying channel propagation. In this section, we introduce a global coordinate system (GCS) for characterizing wireless propagation independent of antenna configuration, and a local coordinate system (LCS) for antenna-specific parameters and beamforming. We then present the antenna array model capturing orientation, pattern, and polarization angle, followed by the channel model and received signal model for multi-panel arrays.

	%			\subsection{REFERENCE COORDINATE SYSTEM AND SPHERICAL ANGLE TRANSFORMATION}
	\subsection{Coordinate systems and angular transformation}
	Beam selection agnostic to antenna orientation requires a GCS to define antenna panel orientation and a LCS where beams are spatially characterized with respect to antenna geometry. The GCS provides a fixed frame of reference in the physical medium independent of antenna array orientation, while the LCS is aligned with the antenna geometry. This dual-coordinate approach enables analytical separation of propagation parameters from antenna configuration parameters, which is fundamental to our heterogeneity-agnostic method. In the GCS, we use the Cartesian standard basis $\unitX, \unitY, \unitZ$ and the spherical basis $\unitr(\ele,\azi), \unitTheta(\ele,\azi), \unitPhi(\ele,\azi)$, where the azimuth $\azi$ and elevation $\ele$ angles are measured from the +x-direction to +y-direction and +z-direction to xy-plane. The corresponding basis vectors and attributes in the LCS are denoted with primes, such as $(\ele',\azi')$. The panel orientation in the GCS is defined by a vector $\orientVec = [\alpha, \gamma, \beta]$, whose entries represent rotations around z, y, x axes. Since beam decisions are made in the LCS based on the signal propagation in the AoA/AoD, the global AoA/AoD angles must be transformed to the LCS. We refer readers to Section 7.1 of \cite{3GPP_TR38901} for well-established definitions of the spherical unit vectors, rotation matrices like $\rotComb{}$, and angular transformation. We express the spherical angular transformation from the GCS to the LCS as a function $\GtoL: \Real^{5} \rightarrow \Real^{2}$, given as \cite{3GPP_TR38901}
	\begin{equation}
		(\theta',\phi') = \GtoL. 
	\end{equation}
	This transformation is used to obtain the local spherical angles of a planar wave with a known global AoA/AoD.
	
	\begin{figure*}
		%				\e{-10pt}
		\centering
		\includegraphics[width=0.95\linewidth]{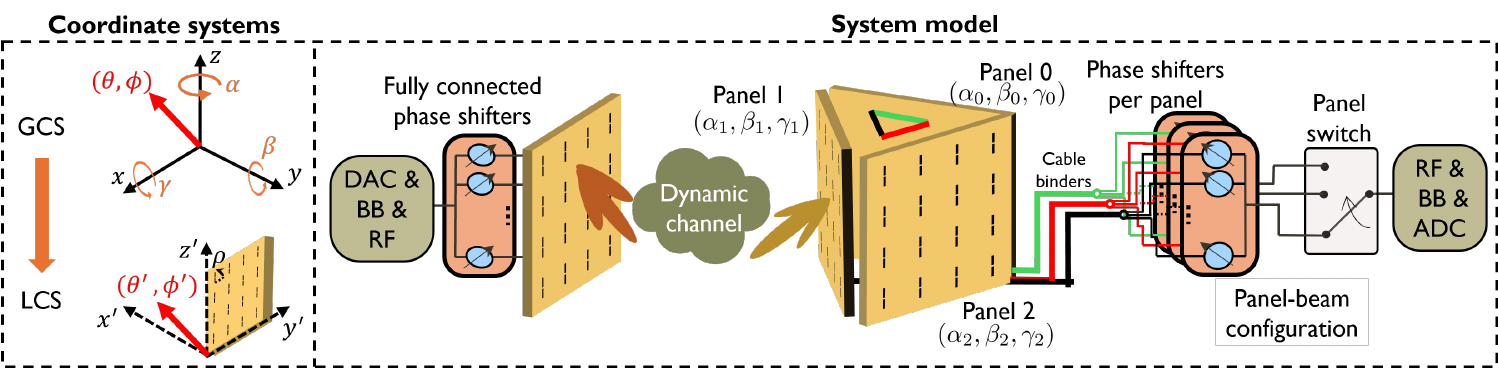}
		%				\vspace{-2pt}
		\caption{Global and local coordinate systems, and the communication system with a BS with single panel and a UE with multi-panel arrays. Antenna panels are UPAs with linearly polarized elements. Each panel has a 3D orientation vector based on the user orientation and the antenna placement. }
		\label{fig:system_model}
	\end{figure*}
	%			\subsection{ANTENNA ARRAY MODEL}
	\subsection{Antenna array model}

	We now develop an antenna array model that analytically incorporates heterogeneity parameters including element patterns, polarization angles, array geometry, and panel orientation. This analytical characterization enables the derivation of a heterogeneity-aware RSRP model in Section \ref{sec:rsrp_agnostic} where these parameters are decoupled from propagation characteristics. We first introduce antenna polarization that refers to the direction of electric field of radiated electromagnetic waves \cite{balanis97}. Antenna elements are designed to have an intended polarization component of the electric field, referred as co-polarization (co-pol). The electric field also has an undesired component perpendicular to the co-polarization, which is known as cross-polarization (cross-pol) \cite{balanis97}. Let $(\ele^\prime,\azi')$ denote the AoA or AoD with respect to the LCS. For a given antenna element, co-pol and cross-pol polarization directions has associated gain patterns denoted as $\cG_\col(\theta',\phi')$ and $\cG_\cross(\theta',\phi')$ \cite{OestgesEtAlMultiPolarizedMIMOCommunicationsChannel2007}. We assume all elements in a panel have the same gain pattern.
	
	Antenna element placement on the antenna array alters co-pol and cross-pol directions. We define the polarization angle $\polAng$ to control the orientation of antenna element on the panel overlaid on the yz-plane of the LCS as shown in \figref{fig:system_model}. Since the co-pol and cross-pol directions are perpendicular to the direction of propagation over $\unitr'(\theta',\phi')$ for the LCS, we can represent them via $\unitTheta'(\theta',\phi')$ and $\unitPhi'(\theta',\phi')$ for the LCS. For notational simplicity, we omit the argument $(\theta',\phi')$ from the unit vectors, which are already defined to be a function of azimuth and elevation angles. The antenna pattern relationship between co/cross-pol directions and $\unitPhi'$/$\unitTheta'$ directions are captured through a rotation matrix $\rotPol \in \mathbb{R}^{2\times2}$ (see the equation (7.1-21) in \cite{3GPP_TR38901}). Let $\mathbf{g}'(\theta',\phi',\polAng) = \trans{\onebytwo{\cG_{\theta'}(\theta',\phi')}{\cG_{\phi'}(\theta',\phi')}}$ denote the gain pattern vector in the $\unitTheta'$ and $\unitPhi'$ directions. This vector is obtained by rotating the co-pol and cross-pol patterns through the polarization angle as $\mathbf{g}'(\theta',\phi',\polAng) = \rotPol \trans{\onebytwo{\cG_\col(\theta',\phi')}{\cG_\cross(\theta',\phi')}}$. The antenna gain pattern $\cG'(\theta',\phi',\polAng)$ in the LCS basis then can be written as $\cG'(\theta',\phi',\polAng) =\onebytwo{\unitTheta'}{\unitPhi'}\mathbf{g}'(\theta',\phi',\polAng)$. It is used to represent the antenna pattern in the GCS basis. 

	The antenna orientation transforms the antenna pattern that is a traversal electric field on the plane perpendicular to the direction of propagation. To represent the wireless propagation independent of the antenna orientation, the LCS gain pattern is represented in the GCS spherical basis. An antenna array with orientation $\orientVec$, and element polarization angle $\polAng$, the pattern transformation matrix $\mathbf{T}(\theta,\phi,\orientVec)$ from local to the global spherical basis can be defined as \cite{3GPP_TR38901}
	\begin{align}
		\bT &=
		\begin{bmatrix}
			\trans{\unitTheta}\rotComb{}\unitTheta'& \trans{\unitTheta}\rotComb{}\unitPhi' \\
			\trans{\unitPhi}\rotComb{}\unitTheta' & \trans{\unitPhi}\rotComb{}\unitPhi'
		\end{bmatrix},
	\end{align} 
	where the spherical angles are omitted for convenience. Let $\mathbf{g}(\theta,\phi,\polAng,\orientVec) =  \trans{\onebytwo{\cG_{\theta}(\theta,\phi)}{\cG_{\phi}(\theta,\phi)}}$ denote the gain pattern vector in the $\unitTheta$ and $\unitPhi$ directions.  This vector is obtained by transforming the gain pattern in the LCS through $\bT$ as $\mathbf{T} \mathbf{g}'(\GtoL,\polAng)$. The global antenna gain pattern $\cG(\theta,\phi,\polAng,\orientVec)$ then is given as $\cG(\theta,\phi,\polAng,\orientVec) = \onebytwo{\unitTheta}{\unitPhi}\mathbf{g}(\theta,\phi,\polAng,\orientVec)$. It analytically captures polarization angle, antenna orientation for a single element pattern.

	Array steering vectors can be described using a vector $\ba(\theta,\phi)$ that characterizes relative phase shifts between antenna elements \cite{heath2018foundations}. Let $d$ be the antenna element spacing in y and z axes, $\lambda$ be the wavelength. For a $N_\text{z} \times N_\text{y}$ UPA in the yz-plane, the steering vector is expressed as
	\begin{equation}
		\begin{split}
			\ba(\theta,\phi) = \trans{\Big[ 
				&1 \cdots 
				e^{-\text{j}2\pi\frac{d}{\lambda}(n_\text{y}\sin\ele\sin\azi + n_\text{z}\cos\ele)}\\
				&\cdots
				e^{-\text{j}2\pi\frac{d}{\lambda}((N_\text{y}-1)\sin\ele\sin\azi + (N_\text{z}-1)\cos\ele)} 
				\Big]},
		\end{split}
	\end{equation}
	where we assumed that the vectorization applied over the y-axis. The array response vectors, though, provides means of analyzing the response arrays with isotropic antennas \cite{BhagavatulaEtAlNewDoubleDirectionalChannelModel2010}. Antenna responses in practice require including the antenna gain patterns and polarizations. We follow the antenna array model in \cite{BhagavatulaEtAlNewDoubleDirectionalChannelModel2010} to model antenna arrays with non-isotropic elements. We assume that each element pattern and the polarization angle in a panel are the same. The combined array response $\bG(\theta,\phi,\polAng,\orientVec) \in \Complex^{N_\text{y}N_\text{z}\times 2}$ is defined as
	%			includes relative phase shifts, polarization angle, antenna orientation and gain pattern in both polarizations
	\begin{equation}\label{eqn:ant_resp_pat}
		\begin{split}
			\bG(\theta,\phi,\polAng,\orientVec) =  \ba(\GtoL) \trans{\mathbf{g}(\theta,\phi,\polAng,\orientVec)} \\
			% = \ba(\GtoL) \mathbf{g}_\text{o}(\GtoL)^\text{T} \rotPol^\text{T}\mathbf{T}(\theta,\phi,\orientVec)^\text{T} 
		\end{split}.
	\end{equation}
	The first and second column in \eqref{eqn:ant_resp_pat} represent array responses in the direction of spherical unit vectors $\unitTheta$ and $\unitPhi$. The combined array response in \eqref{eqn:ant_resp_pat} encapsulates all antenna heterogeneity parameters, i.e., array size through $\ba(\cdot)$, orientation through $\orientVec$, element pattern through $\mathbf{g}(\cdot)$, and polarization through $\polAng$ in an analytical form. This parametrization is crucial in our heterogeneity-agnostic beam selection framework, as it enables explicit separation of these configuration-dependent terms from the propagation-dependent channel characteristics, that is introduced in \secref{subsec:channel_model}.
	
	%			\subsection{CHANNEL MODEL}\label{subsec:channel_model}
	\subsection{Channel model}\label{subsec:channel_model}
	We consider a single-user MIMO (SU-MIMO) communication setting with frequency selective channels, where the BS has a single antenna panel and the UE is a multi-panel antenna array. The system has a static BS and the link is established with one UE panel at a time. The channel model follows double directional clustered channel model capturing antenna array pattern and polarization described in \cite{BhagavatulaEtAlNewDoubleDirectionalChannelModel2010}. It incorporates combined array response with channel depolarization that is a series of scattering processes including specular and diffuse reflections as well as diffraction \cite{BhagavatulaEtAlNewDoubleDirectionalChannelModel2010,HoydisEtAlLearningRadioEnvironmentsDifferentiable2024}. The clustered channel has $\Npath$ paths from the BS to a UE panel. Considering downlink communication, the $\ell$-th path has a complex gain $\pgain$, a path delay $\pdelay$, a angle of departure $(\dod)$ and a angle of arrival $(\doa)$ and a $2 \times2$ depolarization matrix $\pdepol_\ell$, which represents an aggregate matrix for the series of processes on the planar wave over a path \cite{BhagavatulaEtAlNewDoubleDirectionalChannelModel2010,HoydisEtAlLearningRadioEnvironmentsDifferentiable2024}. Let $\pulseshape(\tau)$ be the causal filter that captures low-pass filtering, transmit and receive pulse shaping as a function of $\tau$, and let $T$ denote the symbol period \cite{heath2018foundations}. The BS has a fixed antenna orientation $\orientVec_\tx$ and polarization angle $\polAng_\tx$, and each UE panel $p$ has a different orientation with $\orientVec_{\rx,p}$ but the same polarization angle $\polAng_\rx$. $\bG_{\rx,\ell}(\doa,\polAng_\rx,\orientVec_\rx)$ and $ \bG_{\tx,\ell}(\dod,\polAng_\tx,\orientVec_\tx)$ are the combined array response matrices in \eqref{eqn:ant_resp_pat} for the BS and $p$-th UE panel and we omit the arguments for notational convenience. 
	%equation (12) from linear polarization optimization 
	
	For a channel with $D$ delay taps $\{\bH_d\}_{d=0}^{D-1}$, the discrete delay $d$ channel $\bH_d$ is expressed as \cite{CastellanosHeathLinearPolarizationOptimizationWideband2024,BhagavatulaEtAlNewDoubleDirectionalChannelModel2010}    
	\begin{equation}
		\mathbf{H}_d = \sum_{\ell=1}^{L} \alpha_{\ell} \pulseshape(dT - \tau_{\ell}) \bG_{\rx,\ell}\mathbf{X}_{\ell} \trans{\bG_{\tx,\ell}}.
	\end{equation}
	Let $\text{x}_{\ell,k} = \sum_{d=0}^{D-1} \pulseshape(dT - \tau_{\ell})e^{\frac{-j2\pi kd}{K}}$ and let $\mathbf{X}_{\ell}$ be the $2\times2$ depolarization matrix for the $\ell$-th path. The terms $\pgain$ and $\text{x}_{\ell,k}$ can be combined into a single term $\beta_{\ell,k} = \pgain\text{x}_{\ell,k}$. The frequency response of the channel for the $k$-th subcarrier is then expressed as 
	%equation (14) from linear polarization optimization 
	\begin{equation}\label{eqn:channel_wo_panel}
		\Hk = \sum_{\ell=1}^{L} \beta_{\ell,k} \bG_{\rx,\ell}\mathbf{X}_{\ell} \trans{\bG_{\tx,\ell}}. 
	\end{equation}
	The channel matrix per subcarrier analytically represents antenna heterogeneity parameters and propagation parameters through depolarization matrix, AoAs and AoDs.

	%			\subsection{RECEIVED SIGNAL MODEL}
	\subsection{Received signal model}
	In this work, we consider single user, single stream communication through analog phase shifters both at the BS and the UE panels, and assume perfect synchronization. We assume a single panel operation through a switch circuitry to activate an intended panel in the UE as shown in \figref{fig:system_model}. The analog precoders and combiners at the BS and the UE panels are selected from the BS codebook $\cF$ and the UE panel codebook $\cW_p$ for the $p$-th panel.
	%			 follows industry standard codebook-based configuration \cite{DreifuerstHeathMassiveMIMO5GHow2023}. 
	%			 Let . Let $\bw_p \in \cW_p$ and $\mathbf{f} \in \cF$ denote a combiner at the $p$-th UE panel and precoder at the BS. 
	Let $s[k]$ denote the transmitted symbol with unit power and $P_\tx[k]$ denote the transmit power allocated to the $k$-th subcarrier. We explicitly separate the transmit power from the channel for analytical tractability. In practical systems, the product $\sqrt{P_\tx[k]}\bH_p[k]$ represents the effective channel observed at the receiver. The received signal for the $p$-th panel is then expressed as
	
	\begin{equation}\label{eqn: rec_signal}
		y_p[k] =  \sqrt{P_\tx[k]}\bw_p^* \bH_p[k]\mathbf{f}s[k] + \mathbf{w}^*_\textit{p} \bn[k],
	\end{equation}
	where $\bn[k]$ is the additive noise for the $k$-th subcarrier. \revfive{We use the received signal expression to explore the communication performance of our proposed algorithm, to estimate the effective channel for the RSRP and SNR calculation.}
	
	\section{HETEROGENEITY-AWARE SIGNAL RECEPTION}\label{sec:rsrp_agnostic}
	We develop our heterogeneity-agnostic beam selection based on the RSRP metric per receive and transmit beam as RSRP is used as a standard feedback per synchronization signal block (SSB) beam in 5G \cite{DreifuerstHeathMassiveMIMO5GHow2023,3GPP_TR38901}. Therefore, we start with deriving an RSRP measurement model that preserve analytical decoupling between antenna array configuration, codebooks and propagation characteristics. We later build our heterogeneity-agnostic beam selection method based on a dataset with RSRP measurements. Next, we formulate a path-tracing optimization problem to solve for propagation characteristics, on which we build our heterogeneity-agnostic beam and panel selector. 
	%			\subsection{RSRP DERIVATION}
	\subsection{RSRP derivation}
	By assuming symbols with unit power, instantaneous signal power per subcarrier $\powsig[k]$ using the received signal for $p$-th panel in \eqref{eqn: rec_signal} is given as
	\begin{align}\label{eqn:pow_sig}
		\powsig[k] = P_\tx[k]|\herm{\wpanel} \bH_p[k]\bft|^2  = P_\tx[k]\herm{\wpanel} \Hpanel[k]\bft\mkern1mu\herm{\bft}\herm{\bH}_p[k]\wpanel.
	\end{align}
	The signal power term in \eqref{eqn:pow_sig} captures beamforming, transmit gain and path loss affects. The RSRP is then the average of $\powsig[k]$ over $K$ subcarriers and is given as \cite{ParkParkAnalysisRSRPMeasurementAccuracy2016}
	\begin{align}\label{eqn:rsrp_first}
		\rsrp_\text{exact}  = \frac{1}{K} \sum_{k=1}^{K} P_\tx[k]\herm{\wpanel} \Hpanel[k]\bft\mkern1mu\herm{\bft}\herm{\bH}_p[k]\wpanel.
	\end{align}
	By substituting \eqref{eqn:channel_wo_panel} into \eqref{eqn:rsrp_first} and rearranging summations, the ideal RSRP expression for the panel $p$ expands to
	\begin{align}\label{eqn:rsrp_exact_final}
		\rsrp_\text{exact} &= \frac{1}{K} \sum_{k=1}^{K} P_\tx[k]\herm{\wpanel} \left(\sum_{\ell=1}^{\Npath}\sum_{n=1}^{\Npath} \beta_{\ell,k} \conj{\beta}_{n,k}  \right. \notag \\
		&\quad \left. \times \Grx \bX_{\ell} \trans{\Gtx}\bft\mkern1mu\herm{\bft}\conj{\bG}_{\tx,n}\herm{\bX_{n}} \herm{\bG}_{\rx,n}\right) \wpanel,
	\end{align}
	where orientation vectors, polarization angles, AoAs, AoDs given in \eqref{eqn:ant_resp_pat} are omitted for notational convenience.

	The exact form of the ideal RSRP expression in \eqref{eqn:rsrp_exact_final} includes the cross terms of complex path gains for different paths spatially filtered with receive combiner. In this work, we develop the RSRP model for beam configuration, mainly beam and panel selection problem. In practice, physical channels have many paths, which can be divided into multiple clusters following the clustered channel model. To simplify the RSRP model expression, we assume that the paths in the expression correspond to the dominant paths of different clusters in the channel and dominant paths contribute to the majority of signal power. Therefore, we omit the cross terms in the summation and give the approximate RSRP expression as 
		
	\begin{align}\label{eqn:rsrp_simp}
		\rsrp &= \frac{1}{K} \sum_{k=1}^{K} P_\tx[k]\herm{\wpanel} \left(\sum_{\ell=1}^{\Npath} |\pathgain|^2 \right. \notag \\
		&\quad \left. \times \Grx \bX_{\ell} \trans{\Gtx} \bft\mkern1mu\herm{\bft} \conj{\bG}_{\tx,\ell}\herm{\bX_{\ell}} \herm{\bG}_{\rx,\ell}\right) \wpanel,
	\end{align}
	where only the terms with $\ell = n$ would remain \cite{WallaceJensenSparsePowerAngleSpectrum2009}.
	
	The simplified expression in \eqref{eqn:rsrp_simp} allows us to deal with each path independently and reducing computational complexity. Let $\gamma_{\tx,\ell} = |\trans{\amat}(\GtoLsymb(\theta_{\ell,\tx}, \phi_{\ell,\tx}, \orientVec_{\tx}))\bft|^2$ and $\gamma_{\rx,\ell} = |\herm{\wpanel}\amat(\GtoLsymb(\theta_{\ell,\rx}, \phi_{\ell,\rx}, \orientVec_{\rx}))|^2$. By plugging the combined array response defined in \eqref{eqn:ant_resp_pat} into the RSRP expression, it further simplifies to
	\begin{align}
		\rsrp = \frac{1}{K} \sum_{k=1}^{K}P_\tx[k] \sum_{\ell=1}^{\Npath} |\pathgain|^2 \gamma_{\rx,\ell} \gamma_{\tx,\ell} |\trans{\brx} \bX_{\ell} \btx|^2.
	\end{align}
	In this work, we assume that array response vectors are not frequency dependent. Therefore, the contribution to the RSRP coming from the BS and the UE beams becomes frequency independent. By reordering the summation, the RSRP expression is given as 
	\begin{align}
		\rsrp = \sum_{\ell=1}^{\Npath} \gamma_{\rx,\ell} \gamma_{\tx,\ell} \left(\frac{1}{K} \sum_{k=1}^{K} P_\tx[k]|\pathgain|^2 |\trans{\brx} \bX_{\ell} \btx|^2 \right).
	\end{align}
	Let $\bvec_{\ell} = \brx \otimes \btx$ and $\xvec_{\ell} = \text{vec}(\bX_{\ell})$.
	% \begin{align}\label{eqn:derive_bvec}
		% \bvec_{\ell} &= \brx \otimes \btx\\
		% \xvec_{\ell,k} &= \text{vec}(\bX_{\ell,k}),
		% \end{align}
	$|\trans{\brx} \bX_{\ell} \btx|^2$ then can be written as $|\trans{\bvec_{\ell}} \xvec_{\ell}|^2$. Note that the term within $|\cdot|$ is a complex scalar. We then can write it as $|\trans{\bvec_{\ell}} \xvec_{\ell}|^2 = (\trans{\bvec_{\ell}} \xvec_{\ell})\herm{(\trans{\bvec_{\ell}} \xvec_{\ell})} = \trans{\bvec_{\ell}}\xvec_{\ell} \herm{\xvec}_{\ell} \conj{\bvec}_{\ell}.$
	% \begin{align}
		% |\trans{\bvec_{\ell}} \xvec_{\ell,k}|^2 &= (\trans{\bvec_{\ell}} \xvec_{\ell,k})\herm{(\trans{\bvec_{\ell}} \xvec_{\ell,k})} = \trans{\bvec_{\ell}}\xvec_{\ell,k} \herm{\xvec}_{\ell,k} \conj{\bvec}_{\ell}.
		% \end{align}
	%Let $\bA_{\ell} = \xvec_{\ell,k}\herm{\xvec_{\ell,k}}$,
	The RSRP expression is written as 
	\begin{align}
		\rsrp = \sum_{\ell=1}^{\Npath} \gamma_{\rx,\ell} \gamma_{\tx,\ell} \trans{\bvec_{\ell}}\left(\frac{1}{K} \sum_{k=1}^{K} P_\tx[k]|\pathgain|^2 \xvec_{\ell} \herm{\xvec}_{\ell} \right)\conj{\bvec}_{\ell}.
	\end{align}
	%			Note that $\frac{1}{K} \sum_{k=1}^{K} P_\tx[k]|\pathgain|^2 \Rmatrix_{\ell,k}$ captures channel depolarization, complex path gain values, OFDM subcarriers, transmit power per subcarrier and filtering effects, which we do not need to solve separately for the RSRP estimation when these values are fixed. Usually, the filtering aspects and the number of subcarriers for a communication system is determined in the standards and different devices follow the same convention. We then simplify the summation as $\Rmatrix_{\ell} = \frac{1}{K} \sum_{k=1}^{K} P_\tx[k]|\pathgain|^2 \Rmatrix_{\ell,k}$ and reduce the dimension across subcarriers. Each $\Rmatrix_{\ell,k}$ is an Hermitian, rank-1 matrix by definition. Thus, $\Rmatrix_{\ell}$ is a positive semi definite (PSD), Hermitian matrix \cite{alma9914825166506531} for Hermitian matrix since $P_\tx[k], |\pathgain|^2$ are non-negative coefficients. 
	Let $ \Rmatrix_{\ell}$ be $\frac{1}{K} \sum_{k=1}^{K} P_\tx[k]|\pathgain|^2 \xvec_{\ell} \herm{\xvec}_{\ell}$, where $\xvec_{\ell} \herm{\xvec}_{\ell}$ is a Hermitian, rank-1 matrix. Thus, $\Rmatrix_{\ell}$ is a positive semi definite (PSD), Hermitian matrix \cite{alma9914825166506531} since $P_\tx[k], |\pathgain|^2$ are non-negative coefficients. The final compact ideal RSRP expression is given as
	\begin{align}\label{eqn:ideal_rsrp_final}
		\rsrp = \sum_{\ell=1}^{\Npath} \gamma_{\rx,\ell} \gamma_{\tx,\ell} \trans{\bvec_{\ell}} \Rmatrix_{\ell} \conj{\bvec}_{\ell}.
	\end{align}
	
	The signal measurements in reality are noisy and the ideal RSRP model in \eqref{eqn:ideal_rsrp_final} require an additional term for the noise combined with the receive combiner $\wpanel$. Since the AoA/AoDs and path matrices for $\Npath$ paths are independent of the noise term, the noise power can be expressed as additive. For the IID additive noise, the noise power is defined as $P_\text{n} = \frac{1}{K} \sum_{k=1}^{K}|\wpanel\bn[k]|^2$. Since $P_\text{n}$ is a scalar and it is additive, it can be rewritten as $P_\text{n}=\sum_{\ell=1}^{\Npath} \gamma_{\rx,\ell} \gamma_{\tx,\ell} \trans{\bvec_{\ell}} \tilde{\Rmatrix}_{\ell} \conj{\bvec}_{\ell}$, where $\tilde{\Rmatrix}_\ell$ controls the noise contribution, and other terms are defined based on the AoAs, AoDs, TX-RX beams, antenna orientation, element pattern and polarization angle. Note that $\trans{\bvec_{\ell}} \tilde{\Rmatrix}_{\ell} \conj{\bvec}_{\ell}$ is a scalar and is equal to $ \trace{\trans{\bvec_{\ell}} \tilde{\Rmatrix}_{\ell} \conj{\bvec}_{\ell}} = \trace{\tilde{\Rmatrix}_\ell\bB_\ell}$, where $\bB_\ell=\conj{\bvec}_{\ell}\trans{\bvec_{\ell}}$. The noisy approximate RSRP model is then expressed as 
	\begin{align}\label{eqn:rsrp_final}
		\rsrp_\text{n} = \rsrp + P_\text{n}
		= \sum_{\ell=1}^{\Npath} \gamma_{\rx,\ell} \gamma_{\tx,\ell}\trace{ \hat{\Rmatrix}_\ell\bB_\ell},
	\end{align}
	where $\hat{\Rmatrix}_\ell = \Rmatrix_{\ell}+\tilde{\Rmatrix}_\ell$. Note that $\tilde{\Rmatrix}_\ell$ can be chosen as a PSD matrix for the noise power contribution, then $\hat{\Rmatrix}_\ell$ is also an Hermitian, PSD matrix. \revfive{We use the RSRP model \eqref{eqn:rsrp_final} in the following subsections.}
	
	\begin{algorithm}[htbp]\label{alg:path_tracing}
		\caption{Alternating Path Tracing Optimization}
		\label{alg:ao_rsrp}
		\begin{algorithmic}[1]
			\REQUIRE Max-normalized measurements $\measset$
			\ENSURE Optimized parameters $\angleset^*$, $\matrixset^*$
			
			\STATE Initialize the scaling and iteration index: $\rho = \Nmeas\Npathall$ and $j = 0$
			\STATE Set the max iteration and loss threshold: $j_\text{max}$ and $\epsilon$
			\STATE Initialize $\angleset^{(0)}$: elevation and azimuth angles with $\ele \sim \text{Unif}(0,\pi)$ and $\azi \sim \text{Unif}(-\pi,\pi)$
			\STATE Initialize $\matrixset^{(0)}$: $\hat{\Rmatrix}_\ell^{(0)} = \frac{1}{\rho}\mathbf{I}_4$
			
			\REPEAT
			\STATE $j \leftarrow j + 1$
			
			\STATE \textbf{Step 1: Angle optimization (Powell's Method)}
			\STATE Solve: $\angleset^{(j)} = \arg\min_{\angleset} \optfuncname(\angleset, \matrixset^{(j-1)})$
			\COMMENT{Powell's derivative-free optimization uses $\angleset^{(j-1)}$ as initialization}
			
			\STATE \textbf{Step 2: Joint $\matrixset$ optimization (SDP)}
			\STATE Solve joint SDP:
			\[
			\begin{aligned}
				\matrixset^{(j)} = \argmin_{\{\hat{\Rmatrix}_\ell\}_{\ell=1}^{\Npathall}} &\ \optfuncname(\angleset^{(j)}, \{\hat{\Rmatrix}_\ell\}_{\ell=1}^{\Npathall}) \\
				\text{s.t.} \quad & \hat{\Rmatrix}_\ell \succeq 0, \\
				& \hat{\Rmatrix}_\ell = \herm{\hat{\Rmatrix}_\ell}.
			\end{aligned}
			\]
			
			\STATE Compute objective: $\optfuncname^{(j)} = f(\angleset^{(j)}, \matrixset^{(j)})$
			
			\UNTIL{$f^{(j)} < \epsilon  \textbf{ or }  j \geq j_{\max}$} %$$ 
			
			\RETURN $\angleset^{(j)}$, $\matrixset^{(j)}$
		\end{algorithmic}
	\end{algorithm}
	
	The RSRP measurement model in \eqref{eqn:rsrp_final} analytically captures AoA and AoDs, channel depolarization and path gain in $\hat{\Rmatrix}_\ell$ for each path, the antenna array gains for the beamformer and combiner, and the orientation, polarization and antenna pattern captured in $\bvec_{\ell}$ for each path $\ell$. The number of paths $\Npath$ represents paths from the BS antenna array to a UE panel. All UE panels share the same incoming paths because the panels are closely spaced as shown in \figref{fig:system_model}. Let the total number of paths incoming to all UE panels be denoted by $\Npathall$. The total paths represent the channel between BS antenna and all UE panels, as all panels might not see all paths due to different orientations. Let the sets
	\begin{align}
		\angleset &= \{(\dod,\doa)\}_{\ell=1}^{\Npathall} \quad \text{(AoA/AoD set)} \\
		\matrixset &= \{\hat{\Rmatrix}_\ell\}_{\ell=1}^{\Npathall} \quad \text{(Path matrix set)}  
	\end{align}
	represent \textit{path information} for the wireless propagation agnostic to the antenna heterogeneity. Measured RSRP in practice does not unveil the information of AoA/AoD and path matrix, which we propose solving through optimization using RSRP measurement data with UE locations, panel orientations, polarization angles, antenna patterns, and BS, UE codebooks. We hypothesize that optimized path information can be used to calculate RSRP for beam and panel selection of any antenna configuration. Path information optimization extracts only propagation dependent variables unlike RSRP measurements depending on many variables.
	
	%			\subsection{SOLVING FOR PROPAGATION PARAMETERS}\label{sec:tracing_path_info}
	\subsection{Solving for propagation parameters}\label{sec:tracing_path_info}
	
	We propose an optimization algorithm to solve for path information from RSRP measurements. Let $\measset = \{\rsrp_m\}_{m=1}^{\Nmeas}$ denote the measurement set for a single UE with multi-panel antennas, where each $\rsrp_m$ corresponds to a specific combination of antenna configurations (panel index, orientation, pattern, polarization angle, and beamformer/combiner). \revfive{In our simulations, each measurement is a unique combination. In general, the method does not require exhaustive sampling of all possible configurations. The measurement set must include sufficient diversity across panels and beams to sense distinct spatial directions with varied beam pairs. The required number of measurements is discussed in relation to the problem dimensionality following the optimization formulation}. We assume these configurations are known and characterized by index $m$. Let $\text{RSRP}_m(\angleset,\matrixset)$ denote the RSRP model in \eqref{eqn:rsrp_final}. The path-tracing problem with objective $\optfuncname(\angleset,\matrixset)$ has the form of an inverse problem with the forward generating function $\text{RSRP}_m(\angleset, \matrixset)$ and it is formulated as
	\begin{align}\label{eqn:opt_problem}
		\min_{\angleset, \matrixset} \quad &\frac{1}{\Nmeas} \sum_{m=1}^{\Nmeas} \left|\rsrp_m - \text{RSRP}_m(\angleset, \matrixset)\right|^2 \\
		\text{subject to} \quad &\hat{\Rmatrix}_\ell \succeq 0, \quad \hat{\Rmatrix}_\ell = \herm{\hat{\Rmatrix}_\ell}, \quad \ell = 1, \ldots, \Npathall. \notag
	\end{align}
	
	This problem has no closed-form solution nor a known joint optimization solution, but its structure enables efficient alternating optimization. For a given $\angleset$, optimizing $\matrixset$ reduces to a semidefinite program (SDP) \cite{alma9914825166506531}, which we solve using the splitting conic solver (SCS) \cite{scs} in CVXPY \cite{cvxpy}. Given $\matrixset$, optimizing $\angleset$ is highly non-linear due to coupled trigonometric terms in the combined array response. Therefore, we employ Powell's method \cite{brent2002minimization}, a gradient-free conjugate direction search with automatic step-size control, implemented via \cite{2020SciPy-NMeth}. We do not impose any angular constraints as $\GtoL$ function inherently handles periodicity and projects angles to the desired range.
	
	The Algo. \ref{alg:ao_rsrp} illustrates the alternating path-tracing optimization procedure. RSRP measurements are normalized by the maximum RSRP among all measurements for numerical stability. The algorithm initializes the AoA/AoD set $\angleset$ sampled from uniform distribution, and $\matrixset$ as $\hat{\Rmatrix}_\ell^{(0)} = \frac{1}{\Nmeas\Npathall}\mathbf{I}_4$ to match measurement scales. Each iteration first optimizes $\angleset$ via Powell's method, then solves the SDP for $\matrixset$. The algorithm terminates when the MSE threshold is met or maximum iterations reached. Due to the non-linear objective function, the number of measurements should satisfy $\Nmeas > 36\Npathall$ (4 angles plus 32 real entries per $4\times4$ complex Hermitian matrix, per path). The algorithm solves for the path information which we build our heterogeneity-agnostic beam and panel selection method in the next section.

	\section{HETEROGENEITY-AGNOSTIC BEAM SELECTION}\label{sec:het_agnostic_section}
	
	\begin{figure*}[!ht]
		\centering
		\includegraphics[width=0.95\linewidth]{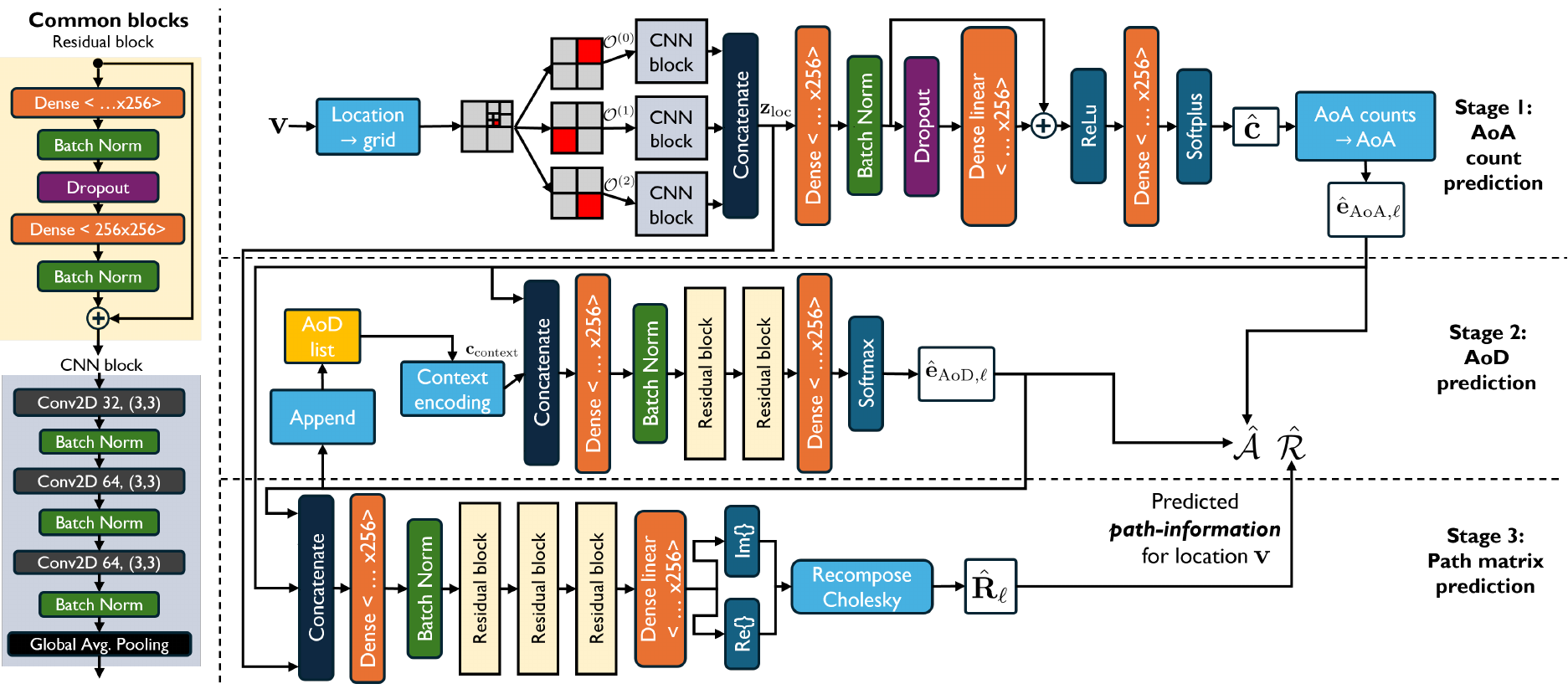}
		\caption{Location-based three-stage path information predictor model. There are two common building blocks shown on the left. Dense layers with no activation specified have ReLU activation. The first stage predicts the AoA counts per angle cluster given the location information. Stage 2 performs an autoregressive AoD prediction for each AoA cluster index so that the context information avoids predicting the same AoD indices for the AoA index with more than one occurrence. Finally, stage 3 performs path matrix prediction given AoA, AoD one-hot vectors per path and the location.}
		\label{fig:path_predictor}
	\end{figure*}

		In this section, we explain the position-aided heterogeneity agnostic beam selection method for multi-panel antenna arrays. Our methodology leverages that path information $(\angleset, \matrixset)$ is inherently independent of antenna heterogeneity, as established in the RSRP model of \secref{sec:rsrp_agnostic}. Given path information and antenna configuration parameters, RSRP values can be calculated analytically for any beam-panel combination regardless of antenna size, orientation, polarization, or codebook. This enables our heterogeneity-agnostic approach. Instead of directly predicting RSRP values for specific antenna configurations, underlying path information is predicted based on UE locations, which is then used to compute RSRP for arbitrary heterogeneous configurations. We first develop a three-stage neural network that maps UE locations to path information obtained via the optimization framework in \secref{sec:tracing_path_info}. In the model deployment, the predicted path information $(\hat{\angleset}, \hat{\matrixset})$ is then combined with the known antenna heterogeneity parameters to analytically calculate predicted RSRP values $\widehat{\rsrp}_{p,i}$ for each beam $i$ and panel $p$. This approach enables beam and panel selection across diverse antenna configurations without any retraining.

		\subsection{Het-RSRPredictor} \label{sec:het_rsrp_predictor}

		\textit{Het-RSRPredictor} is a heterogeneity-agnostic RSRP predictor model based on UE locations. Het-RSRPredictor has a three-stage neural network that predicts path-information $(\angleset, \matrixset)$ for a given user location $\xloc$ as shown in \figref{fig:path_predictor}. We assume that a user can obtain position through inertial sensors or GPS, which most modern UE devices support \cite{Gonzalez-PrelcicEtAlIntegratedSensingCommunicationRevolution2024}. RSRP is a readily accessible metric in 5G \cite{DreifuerstHeathMassiveMIMO5GHow2023,3GPP_TR38901} and can be collected and updated systematically to generate datasets to train and fine-tune AI/ML based wireless methods. We assume an RSRP dataset $\{\measset_i\}_{i=1}^{\Nusers}$ containing measurements from $\Nusers$ users, where each $\measset_i$ represents RSRP observations collected from user $i$. Through path-tracing algorithm, each $\measset_i$ yields corresponding angle sets $\angleset_i$ and path matrix sets $\matrixset_i$. The complete labeled dataset is then formulated as $\cD = \{(\xloc_i, \angleset_i, \matrixset_i)\}_{i=1}^{\Nusers}$, where $\xloc_i \in \mathbb{R}^2 = \trans{[\text{x},\text{y}]}$ denotes user coordinates in the GCS. 
		
		% Paragraph 1: Overview + K-means
		Predicting path-information is challenging due to the permutation invariance of paths, where different orderings of angle sets represent identical RSRP measurements. A physically consistent method of organizing path predictions is required to handle this challenge. We develop a novel three-stage path predictor that combines angle quantization through spatial clustering with hierarchical grid-based location encoding. The continuous angle space is first discretized using Euclidean K-means clustering \cite{2020SciPy-NMeth} applied to unit vector representations in the Cartesian basis. Each angle pair $(\theta, \phi)$ is converted to a 3D unit vector $\uvec \in \Real^3$ on the unit sphere. The K-means algorithm partitions these unit vectors into $\Ngrid$ clusters $\{C_1, C_2, \ldots, C_{\Ngrid}\}$, where $\boldsymbol{\mu}_g$ represents the center of cluster $g$. \revfive{Since all vectors have unit norm, the Euclidean K-means clustering naturally preserves the spherical manifold structure}. Each angle is subsequently assigned to its nearest cluster and represented as a one-hot vector $\onehot{g} \in \{0,1\}^{\Ngrid}$ where the index $g$ indicates the cluster assignment. This preserves directional relationships through the spherical geometry while enabling tractable prediction over a finite angle set. To extract representative features in higher dimensions, 2D locations are encoded with a hierarchical grid representation that provides multi-scale spatial features from coarse to fine grid granularity. This representation is particularly effective for CNN-based \cite{goodfellow2016deep} feature extraction.
		
		% Paragraph 2: Hierarchical grid encoding
		\revfive{The hierarchical grid encoder transforms UE locations into multi-level spatial representations. The encoder operates at $\Mgrid$ hierarchical levels, with each level using a spatial grid of size $\Gs \times \Gs$. For each UE location $\xloc_i = \trans{[x_i, y_i]}$, the encoding function $\gridenc(\xloc): \Real^2 \rightarrow \{0,1\}^{\Mgrid \times \Gs \times \Gs}$ produces a multi-level spatial representation $\gridtensor_i \in \{0,1\}^{\Mgrid \times \Gs \times \Gs}$}. The encoding process begins by defining a bounding rectangle on the xy-plane that covers all possible UE locations and dividing it into a $\Gs \times \Gs$ grid uniform within each axis. For each hierarchical level $\gridindex = 0, \ldots, \Mgrid-1$, the encoder identifies the corresponding grid cell and activates it by setting $[\gridlevel{\gridindex}]_{o,f} = 1$ for the appropriate cell $(o,f)$. Level $\gridindex=0$ represents the coarsest granularity, dividing the entire area into $\Gs \times \Gs$ cells. Each subsequent level progressively zooms into the activated cell from the previous level, subdividing it into $\Gs \times \Gs$ finer cells. This creates a pyramid of increasingly precise spatial information, where level $\gridindex$ provides location resolution at scale proportional to $(1/\Gs)^{\gridindex+1}$. The hierarchical structure provides an efficient multi-scale encoding where each location activates exactly one cell per level, and avoids modeling all possible grid positions.

		% Paragraph 3: CNN grid processing
		We develop a convolutional grid processor neural network architecture that extracts spatial features from hierarchical grids independently before combining them. This grid processor $f_{\text{grid},\boldsymbol{\theta}_{\text{grid}}}: \{0,1\}^{\Mgrid \times \Gs \times \Gs} \rightarrow \Real^{\dgrid}$ is shared between Stage 1 and Stage 3 of path information predictor to leverage spatial location information. For each hierarchical level $\gridindex$, the corresponding grid $\gridlevel{\gridindex}$ is processed through three convolutional layers, each followed by batch normalization. The convolutional operations with ReLU activations enable the network to learn spatial patterns and relationships within each grid level \cite{goodfellow2016deep}. After the third convolutional layer, global average pooling reduces each level's feature map to a fixed-size vector $\hlevel{\gridindex} \in \Real^{64}$. The per-level features from all $\Mgrid$ levels are then concatenated to produce the final location feature vector $\hloc = [\hlevel{0} \| \hlevel{1} \| \cdots \| \hlevel{\Mgrid-1}] \in \Real^{\dgrid}$, where $\|$ denotes concatenation \cite{goodfellow2016deep}. This architecture enables the network to simultaneously produce both coarse-scale and fine-scale location feature information.
		
		Stage 1 implements AoA count prediction by mapping hierarchical location grids to a count vector $\cvec_i \in \Real_+^{\Ngrid}$ that specifies the predicted number of paths for each AoA cluster. Since multiple propagation paths may arrive from similar directions and thus belong to the same AoA cluster $g$, the count $c_{i,g}$ represents how many of the $\Npathall$ total paths are associated with that cluster. This count-based formulation naturally handles the permutation invariance of paths within each cluster while maintaining the total number of paths as $\sum_{g=1}^{\Ngrid} c_{i,g} = \Npathall$. The AoA counts are converted to a set of AoA indices for the Stage 2 AoD prediction. The network processes the input grids $\gridtensor_i$ through the CNN grid processor to obtain location features $\hloc$, which are then passed through dense layers parameterized by $\boldsymbol{\theta}_1$ with a residual connection. The output layer has softplus activation \cite{goodfellow2016deep} to ensure non-negative count predictions. The training employs a combined loss function balancing multiple objectives. The KL divergence term enforces similarity between true and predicted count distribution, given for a single sample as \cite{goodfellow2016deep}
		\begin{align}
			\Lkl(\boldsymbol{\theta}_{\text{grid}}, \boldsymbol{\theta}_1) = \sum_{g=1}^{\Ngrid} \tilde{c}_{i,g} \log\frac{\tilde{c}_{i,g}}{\tilde{\hat{c}}_{i,g}},
		\end{align}
		where $\tilde{c}_{i,g} = c_{i,g}/\sum_{g'} c_{i,g'}$ and $\tilde{\hat{c}}_{i,g} = \hat{c}_{i,g}/\sum_{g'} \hat{c}_{i,g'}$ are the normalized count distributions. The total count preservation loss ensures the predicted AoA counts sum to approximately $\Npathall$ paths. It is given as 
		\begin{align}
			\Ltotal(\boldsymbol{\theta}_{\text{grid}}, \boldsymbol{\theta}_1) = \left(\sum_{g=1}^{\Ngrid} c_{i,g} - \sum_{g=1}^{\Ngrid} \hat{c}_{i,g}\right)^2.
		\end{align}
		The sparsity regularization encourages concentrated predictions by penalizing counts above threshold $\taus$, given as 
		\begin{align}
			\Lsparse(\boldsymbol{\theta}_{\text{grid}}, \boldsymbol{\theta}_1) = \frac{1}{\Ngrid}\sum_{g=1}^{\Ngrid} \text{ReLU}(\hat{c}_{i,g} - \taus).
		\end{align}
		The complete Stage 1 loss combines mean absolute error with these losses and it is defined as
		\begin{align}
			\Lcount(\boldsymbol{\theta}_{\text{grid}}, \boldsymbol{\theta}_1) = \frac{1}{\Ngrid}\|\cvec_i - \hat{\cvec}_i\|_1 \notag \\
			\quad + \lambdakl \Lkl + \lambdatotal \Ltotal + \lambdasparse \Lsparse,
		\end{align}
		where each loss term has an associated weight $\lambda$ to determine importance on regularization \cite{goodfellow2016deep}. 
		
		% Paragraph 5: Stage 2 with lightweight context
		Stage 2 predicts AoD indices autoregressively given the AoA cluster through the function $f_{2,\boldsymbol{\theta}_2}(\onehot{g}, \ctx) \rightarrow \hat{\pvec}_{\text{AoD}} \in [0,1]^{\Ngrid}$, where $\onehot{g}$ is the one-hot encoded AoA cluster vector and $\ctx \in \Real^{\dcontext}$ is a context vector. The AoA one-hot vector and context are concatenated and processed through dense layers with two residual blocks. A softmax output layer produces a probability distribution over AoD clusters from which the AoD index $g_{\text{AoD}}$ is selected. The context mechanism tracks previously predicted AoDs for the current AoA cluster. For each AoA cluster $g$, the context is initialized as $\ctx = \mathbf{0} \in \Real^{\dcontext}$ before the first prediction. After each AoD prediction with index $g_{\text{AoD}}$, the context is updated by incrementing $\ctx[g_{\text{AoD}} \bmod \dcontext]$ by one. The modulo operation maps AoD cluster indices to the context dimension. This accumulator mechanism biases the network toward unexplored directions and promotes diversity in predictions. The training is done using categorical cross-entropy loss \cite{goodfellow2016deep} and it is defined as 
		\begin{align}
			\Ltwo(\boldsymbol{\theta}_2)= -\sum_{g=1}^{\Ngrid} e_{\text{AoD},i,\ell,g} \log(\hat{p}_{i,\ell,g}),
		\end{align}
		where $\mathbf{e}_{\text{AoD},i,\ell}$ is the true AoD vector for sample $i$ and path $\ell$, and $\hat{p}_{i,\ell,g}$ is the predicted probability for cluster $g$.

		% Paragraph 6: Stage 3 with magnitude-aware loss
		Stage 3 predicts path matrices through the function $f_{3,\boldsymbol{\theta}_3}(\hloc, \onehot{\text{AoA},i,\ell}, \onehot{\text{AoD},i,\ell}) \rightarrow \widehat{\text{vec}(\Lvec_{\ell,i})} \in \Real^{20}$, which maps the location features and angle information to the path matrix representation. The location features $\hloc$ extracted by the CNN grid processor in the Stage 1 are concatenated with the one-hot encoded AoA and AoD vectors. The concatenated features pass through dense layers parameterized by $\boldsymbol{\theta}_3$ with three residual blocks to capture complex relationships between location, angles, and channel characteristics. The residual connections maintain gradient flow through the deep network while dropout regularization prevents overfitting \cite{goodfellow2016deep}. The output layer produces a real-valued vector that is reshaped and converted to the complex Cholesky factor $\hat{\Lvec}_{i,\ell} \in \Complex^{4 \times 4}$ \cite{alma9914825166506531}. The path matrix is then reconstructed via Cholesky decomposition as $\hat{\Rmatrix}_{i,\ell} = \hat{\Lvec}_{i,\ell} \hat{\Lvec}_{i,\ell}^*$, which ensures the resulting matrix is Hermitian and positive semidefinite without requiring explicit constraints \cite{alma9914825166506531}. In the training, we define a magnitude-aware MSE loss that prioritizes accurate prediction of path matrix magnitudes. The magnitude error is computed as
		\begin{align}
			\Lmag(\boldsymbol{\theta}_3) = \frac{1}{10}\|\,|\text{vec}(\Lvec_{\ell,i})| - |\widehat{\text{vec}(\Lvec_{\ell,i})}|\,\|_2^2,
		\end{align}
		The direction error measures the reconstruction performance, given as
		\begin{align}
			\Ldir(\boldsymbol{\theta}_3) = \frac{1}{20}\|\text{vec}(\Lvec_{\ell,i}) - \widehat{\text{vec}(\Lvec_{\ell,i})}\|_2^2.
		\end{align}
		The final Stage 3 loss combines these two terms as
		\begin{align}
			\Lthree(\boldsymbol{\theta}_3) = \Ldir + \lambdamag \Lmag.
		\end{align}
		This loss formulation ensures the reconstructed path matrices have appropriate energy characteristics for accurate RSRP prediction across heterogeneous antenna configurations.

		%	\subsection{Static Pred: STATIC BEAM-PANEL SELECTOR}\label{sec:static_pred}
		\subsection{Static Pred: Static beam-panel selector}\label{sec:static_pred}
		
		The static beam selector \textit{Static Pred} relies on RSRP prediction values from Het-RSRPredictor. We make several practical assumptions for beam-panel selection at the UE. First, we assume the BS heterogeneity is known and readily available at the UE, which can be transferred with Het-RSRPredictor. Second, based on our prior work \cite{KilincEtAlBeamTrainingMmWaveVehicular}, where BS beam configuration can be decoupled from the UE with minimal performance loss, and \cite{AliEtAlOrientationAssistedBeamManagement5G2021} we assume the optimal BS beam index is known at the UE. These assumptions are not required to unlock the full potential of our proposed solution but provide clear insights of heterogeneity-agnostic beam-panel selection for multi-panel arrays. The training, fine-tuning, and deployment in specific applications can vary with different assumptions, which we leave as a future work.
		
		The beam-panel selection operation flows from UE location to RSRP calculation to subset selection. Given a UE location $\xloc$, the path-predictor network in \figref{fig:path_predictor} predicts the path information $\hat{\angleset},\hat{\matrixset}$, which is then used to calculate RSRP predictions for all beam-panel combinations given the UE heterogeneity. The predictor network implicitly captures the distributional characteristics of the paths over the locations through the custom loss functions. Let $|\cW_p|$ denote the cardinality of the UE codebook for the $p$-th panel, $i$ denote the beam index where $i = 1\ldots|\cW_p|$, and $\widehat{\rsrp}_{p,i}$ denote the predicted RSRP for the $i$-th beam $\bw_{p,i}$ of the $p$-th panel. Let $\text{topNb}(\cdot)$ be the function returning the set of $\Nbeam$ beams $\bw_{p,i}$ with the highest $\widehat{\rsrp}_{p,i}$ values sorted in descending order. The task is to select a beam-panel subset $\cS$ with cardinality $\Nbeam$, where $\cS \subset \bigcup_p \cW_p$. The selection is given as
		\begin{equation}
			\cS = \text{topNb}\left(\{\widehat{\rsrp}_{p,i}|\forall{p=1,\ldots,P}, i=1,\ldots|\cW_{p}|\}\right).
		\end{equation}
		Beam sweeping over $\cS$ determines the best UE panel and corresponding beam. Keeping $\Nbeam\ge1$ is advantageous in rapid changes of optimal panels and beams.
		
		\section{SIMULATIONS}\label{sec:simulation}
		% \section{Simulation methodology}\label{sec:simulation}
		In this section, we first present simulation setup, the training of the heterogeneity-agnostic beam predictor, and then data transmission protocol. Lastly, we present simulation results of proposed beam selection compared to baselines. 
		%			\subsection{SPATIALLY CONSISTENT CHANNEL GENERATION FOR MOBILE ENVIRONMENT}
		\subsection{Spatially consistent channel generation for mobile environment}
\begin{figure}
			\centering
			\includegraphics[width=0.80\linewidth]{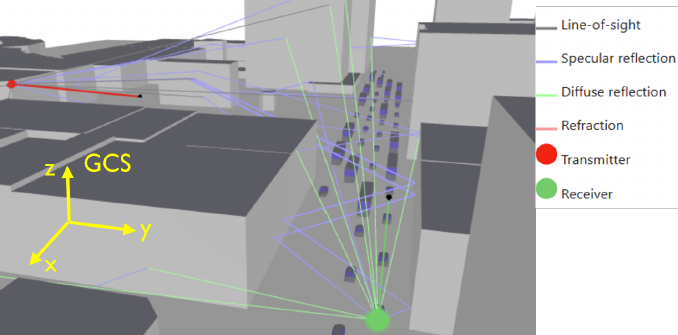}
			\caption{Urban canyon environment with dynamic vehicles. The buses are blockers and cars are users equipped with multi-panel antenna arrays. The solid bold red and green arrows show the orientation of the BS panel and a panel of a UE with multi-panels. Channels for each BS panel and UE panel pairs are generated in Sionna \cite{hoydis2023sionna}. Different types of traced paths is visible in the legend.}
			\label{fig:simulation_setup}
		\end{figure}

		Spatially and physically consistent channel generation is crucial to evaluate beam selection solutions in wireless systems. Ray-tracing is an effective approach for realistic channel modeling that maintains both physical and spatial consistency \cite{RezaieEtAlDeepLearningApproachLocation2022,WangEtAlMmWaveVehicularBeamSelection2019}. We use Sionna \cite{hoydis2023sionna}, an open-source ray-tracing framework that enables flexible channel generation in custom environments. In our simulations, we generate channels in V2I settings, though it is just a simulation choice and our method is not limited to vehicular applications. We generate channels for 10k distinct users for 600 snapshots of a four-way road lane, where vehicles and blockers are initialized randomly as shown in \figref{fig:simulation_setup}. The carrier frequency is set to 15GHz \revfive{that is a candidate frequency for upper mid-band. There are} 64 OFDM subcarriers with 240kHz subcarrier spacing. The noise figure is 10dBm \cite{GraffEtAlDeepLearningBasedLinkConfiguration2023}, thermal noise power density is -174dBm/Hz, and the TX power over the band is 11dBm, calculated based on the available bandwidth \cite{GraffEtAlDeepLearningBasedLinkConfiguration2023}. The BS codebook is a DFT codebook with 64 transmit beams. The UE panel codebooks are angle-limited codebooks constructed by uniformly sampling array steering vectors at $\Nry$ azimuth angles $\phi \in [-60^{\circ}, 60^{\circ}]$ and $\Nrz$ elevation angles $\theta \in [30^{\circ}, 150^{\circ}]$. The choice of angular ranges is based on the multi-panel structure we consider as shown in \figref{fig:system_model} to cover 3D with distinct beams.
		
		The simulation environment consists of a four-way road lane with a length of 300m and bidirectional traffic following US conventions as shown in \figref{fig:simulation_setup}. Mobile vehicles travel with an average speed of 55km/h. A BS equipped with an $8\times8$ UPA is mounted on a building wall approximately 150m from the road center. Each vehicle has multi-panel antennas with $P=3$ panels placed on the roof within the shark fin at the rear. The multi-panel structure illustrated in \figref{fig:system_model} comprises 3 panels arranged such that with the default orientation (Orient $0.0^{\circ}$), panel-0 faces the back of the vehicle while panels 1 and 2 face the front right and left, forming an equilateral triangle configuration.
		
		% - 98 -> 90
		%			\subsection{TRAINING HET-AGNOSTIC BEAM PREDICTOR}
		\subsection{Training het-agnostic beam predictor}
		The distinct users of 10k are split into train, test and validation sets ($70\%$, $20\%$, $10\%$). Since we test the performance under antenna heterogeneity, we generate multiple versions of channel datasets with varying antenna sizes, orientations. The heterogeneity of the BS is fixed, i.e., 3GPP `tr38901' pattern with polarization angle $\polAng_\tx = 0^{\circ}$ (vertical polarization) in \cite{hoydis2023sionna}, $8\times8$ antenna array and a fixed orientation shown by the red arrow in \figref{fig:simulation_setup}. The UE configurations are $3\times3, 5\times5, 7\times7$ for antenna size, $5$ distinct orientations of multi-panel structure along z-axis. Each orientation is characterized by an offset to the default orientation and the offset is expressed as $\Delta\alpha = 24^{\circ}\times a \text{ for }a = 0\ldots4$. The training configuration for beam selection is $3\times3$, $\Delta\alpha=0$ and $\polAng_\rx=0$, the total number of traced paths $\Npathall$ is set to $3$ and the angle grid size $G$ is set to $120$. We use other configurations to test the efficacy of the proposed method. 
		
		The RSRP values for each BS precoder and UE combiners across all panels are estimated with the training configuration for each sample in each dataset. The transmit power, noise figure and thermal noise expressions are reproduced as in \cite{GraffEtAlDeepLearningBasedLinkConfiguration2023} based on the fraction of bandwidth considered in our work. We use least-squared (LS) channel estimation and Zadoff-Chu sequences to estimate beamformed channel in frequency domain to get the RSRP estimate averaged all subcarriers \cite{heath2018foundations}. Next, alternating path-tracing optimization is performed and training RSRP estimates are converted to path information per training sample. The AoA/AoD set is clustered and quantized into AoA/AoD indices through K-means \cite{2020SciPy-NMeth} and the path matrices are decomposed into lower dimensions through Cholesky decomposition \cite{alma9914825166506531}. We independently train each stage of the path predictor in Het-RSRPredictor using the Adam optimizer \cite{goodfellow2016deep}, and with a learning rate of $10^{-3}$, $\lambdakl = 0.3$, $\lambdatotal=0.2$, $\lambdasparse=0.05$, $\lambdamag=2$, $\taus = 0.5$, $\dcontext=60$, and $\Mgrid=3$. The grid processor is only trained at Stage 1, and is frozen.
		
		\begin{figure}[htbp]
			\centering
			\begin{minipage}[b]{0.2\textwidth}
				\centering
				\includegraphics[width=3.7cm]{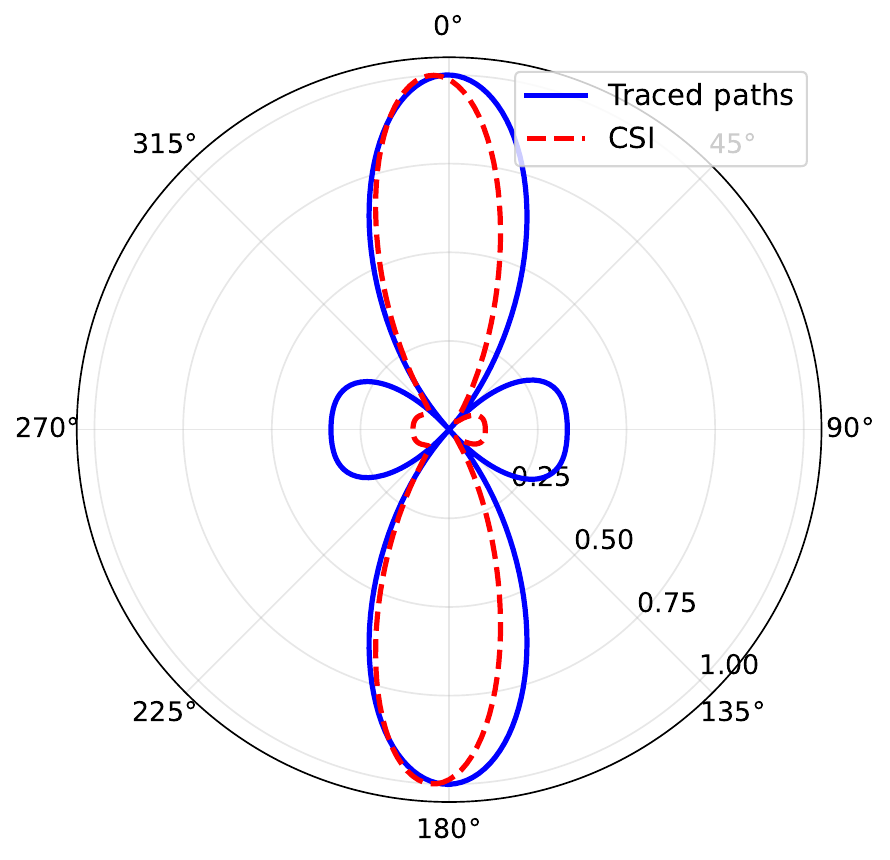}
				\\[0.1em]
				{\small \textbf{(a)} UE panel-1}
			\end{minipage}
			\hfil
			\begin{minipage}[b]{0.2\textwidth}
				\centering
				\includegraphics[width=3.7cm]{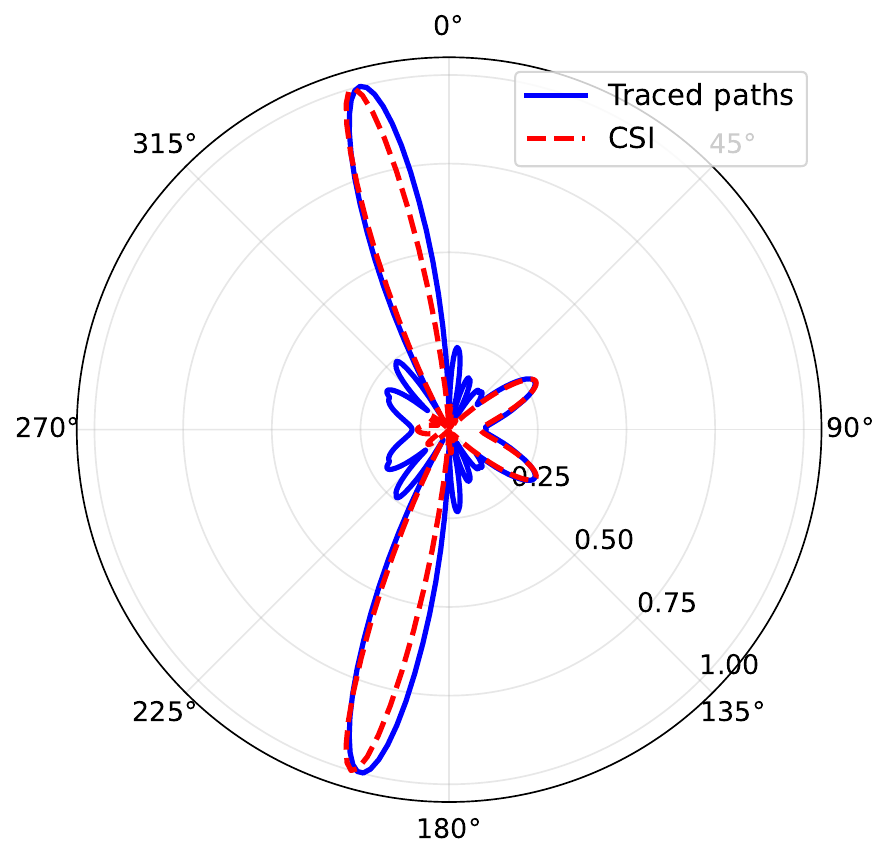}
				\\[0.1em]
				{\small \textbf{(b)} BS antenna panel}
			\end{minipage}
			\caption{Normalized signal power comparison between perfect CSI and traced paths at elevation angle $90^\circ$. In (a), the BS beam is fixed to a DFT codebook beam; in (b), the UE beam is fixed. Traced paths from noisy RSRP provide angular power profiles matching perfect CSI.} 
			\label{fig:user443_panels}
		\end{figure}
		
		%			\subsection{DATA TRAN/SMISSION}
		\subsection{Data transmission}
		Before the data transmission phase on the downlink, the user determines a subset of beam and panel pairs via the proposed beam and panel selection method. Once a subset is determined, a beam sweeping is performed on the selected beam and panel indices. During beam sweeping, the UE groups beams by panel such that all beams from the same panel are tested consecutively before switching to a different panel. This approach minimizes the overhead associated with switching between panels. At the end of beam training, the best beam and panel pair is determined and the data transmission starts \cite{DreifuerstHeathMassiveMIMO5GHow2023}.

		Let $P_\text{n}[k]$ denote the noise power and $h_\text{eff}[k]=\bw_p^* \bH_p[k]\mathbf{f}$ denote the effective channel for $k$-th subcarrier. Let $\hat{h}_\text{eff}[k]$ denote the LS channel estimate \cite{heath2018foundations} for $k$-th subcarrier. We assume total transmit power $P_\tx$ is uniformly distributed across all subcarriers. The effective SNR per subcarrier is then given as
		\begin{equation}
			% SNR calculation for a selected beam
			\text{SNR}_\text{eff}[k] = \frac{P_\tx|\hat{h}_\text{eff}[k]|^2|s[k]|^2}{K P_\text{n}[k]}.
		\end{equation}			
		Let $T_\text{coh}$ denote the beam coherence time, and $T_\text{train}$ denote the beam training overhead, which is defined as $N_\text{ss}T_\text{sym}\Nbeam$, where $N_\text{ss}$ is the number of OFDM symbols, $T_\text{sym}$ is the OFDM symbol duration, and $\Nbeam$ is the selected beam-panel subset cardinality \cite{GraffEtAlDeepLearningBasedLinkConfiguration2023}. With Gaussian noise and signaling and uniform power allocation across subcarriers, the achievable spectral efficiency (SE) over the band is given as \cite{heath2018foundations} 
		\begin{equation}
			% Spectral efficiency calculation across all subcarriers
			\text{SE} = \text{max}\left(0,1-\dfrac{T_\text{train}}{T_\text{coh}}\right)\dfrac{1}{K}\sum_{k=1}^{K} \log_2(1 + \text{SNR}_\text{eff}[k]).
		\end{equation}
		We set $N_\text{SS} = 4$ by following 3GPP NR \cite{GraffEtAlDeepLearningBasedLinkConfiguration2023}. We use achievable SE as the key metric to evaluate the performance of het-agnostic beam and panel selection method. The beam coherence time $T_\text{coh}$ is a parameter that captures how frequently beam training occurs. It varies with user mobility, environment geometry, and channel dynamics. We treat $T_\text{coh}$ as a variable in our analysis to evaluate beam training overhead. While exact characterization of $T_\text{coh}$ for specific environments is beyond our scope, varying this parameter reveals the regimes where the proposed approach is advantageous over conventional methods.

		\subsection{Benchmarks}
		% \subsection{Achieving het-agnostic beam selection}
		In the analysis, we evaluate two variants of the proposed solution and four benchmarks. Our methods are Static Pred explained in \secref{sec:static_pred} and Traced Path, which uses the path information directly from the optimization. The motivation of Traced Path is to determine and an upper bound to  Static Pred. The benchmarks are summarized as follows:
		
		\begin{figure}[ht]
			\centering
			\includegraphics[width=\figsize \linewidth]{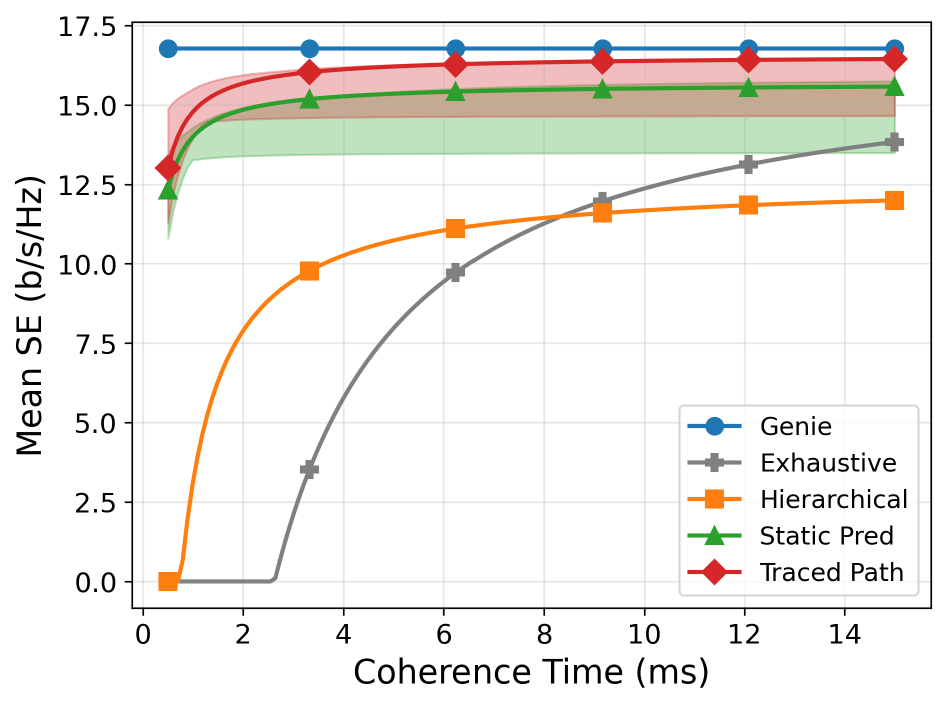}
			\caption{Mean SE vs beam coherence time $T_{\text{coh}}$ for $7\times7$ UE panels. Shaded regions show Pareto frontiers for Static Pred and Traced Path with $\Nbeam$ from 1 to 10 and solid lines use $\Nbeam=6$. Path tracing achieves within 0.5 b/s/Hz of the genie-aided method under antenna size heterogeneity.}
			\label{fig:het_antenna_size}
		\end{figure}

		\textit{Genie}: The true RSRP values are known and the best UE panel and beam is selected based on maximum RSRP. There is no offline training and no beam training overhead.
		
		\textit{Exhaustive}: This is a brute-force sweeping of all possible UE beams across all panels to determine beam-panel with maximum RSRP. The beam training overhead is $\Nry \times \Nrz \times P$ beams varying with respect to the array dimension.

		\textit{Hierarchical}: The hierarchical search is a two-tier beam search. The number of beams at each tier is fixed as the UE panel dimension, i.e., 3 beams for $3\times3$. In first-tier codebooks, the array front over half sphere is clustered into the number of beams to determine the main beam direction and the angular region for each coarse beam by Fibonacci sampling \cite{AliEtAlOrientationAssistedBeamManagement5G2021} and K-means \cite{2020SciPy-NMeth}. The beam directions for the second-tier are obtained in the same way and beams are generated using array steering vectors in the main beam direction. During the beam sweeping, the best beam and panel for the first-tier codebooks are found per panel. In the next stage, the second-tier codebooks for the selected coarse beam per panel are searched and the best beam and panel is determined. There is no offline training and the overhead is $2\times\Nry \times P$ beams varying w.r.t. the array dimension.

		\textit{Baseline}: The baseline is a neural network that predicts the RSRP of the UE beams per panel given the orientation and location of the antenna panel \cite{RezaieEtAlDeepLearningApproachLocation2022}. The beam and panel subset is selected through the highest predicted RSRP values across all beams and panels. The method, however, is not able to perform with different antenna dimensions since it is trained for a fixed antenna dimension.

		\begin{figure}[htbp]
			\centering
			\begin{minipage}[b]{0.48\columnwidth}
				\centering
				\includegraphics[width=\textwidth]{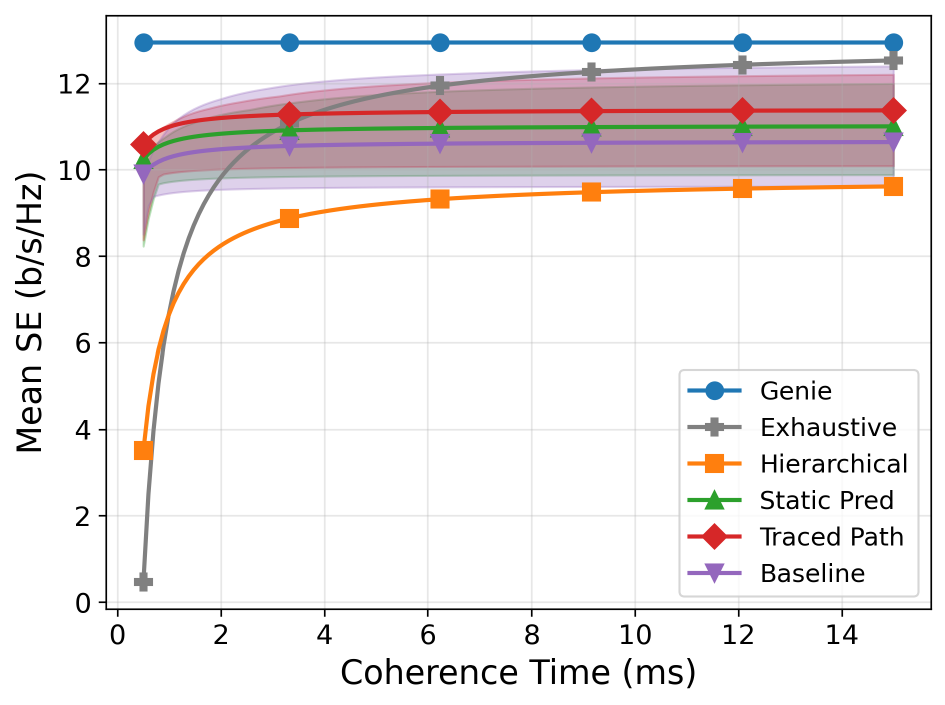}
				\\[0.1em]
				{\small \textbf{(a)} Mean SE vs $\Tcoh$}
			\end{minipage}
			\hfil
			\begin{minipage}[b]{0.48\columnwidth}
				\centering
				\includegraphics[width=\textwidth]{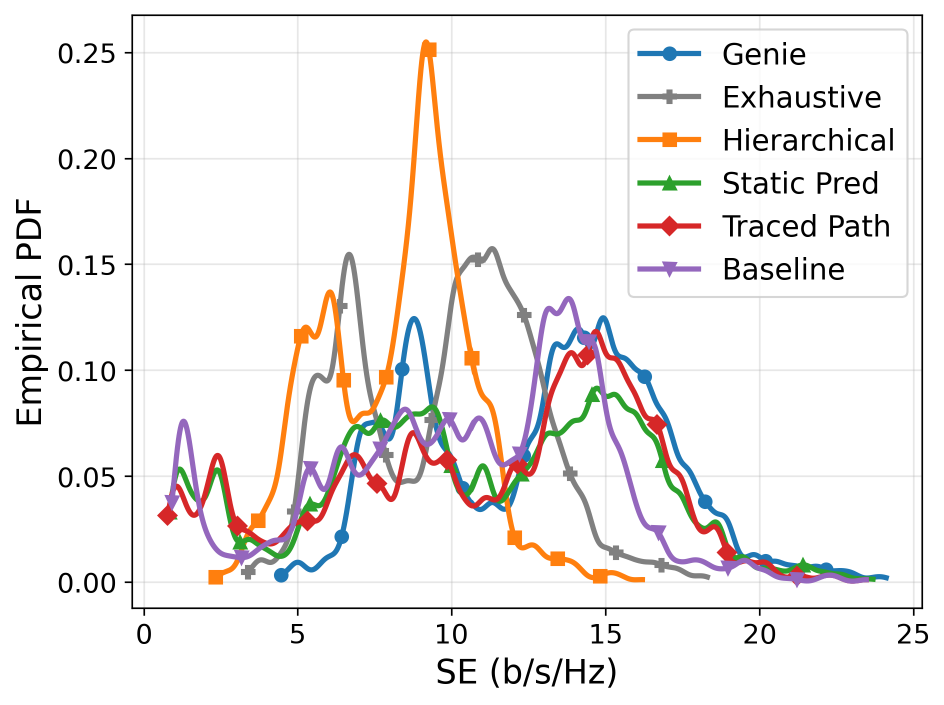}
				\\[0.1em]
				{\small \textbf{(b)} Empirical PDF of SE} 
			\end{minipage}
			\caption{Mean SE and user-for-user SE for orientation heterogeneity with $\Delta\alpha = 96^\circ$ and $3\times3$ UE panels. In (a), Pareto frontiers vary $\Nbeam$ from 1 to 10. In (b), $\Tcoh = 2$ms. Static Pred and Traced Path achieve better user-for-user SE than Baseline under antenna orientation heterogeneity.}
			\label{fig:het_orient}
		\end{figure}

		\subsection{Achieving het-agnostic beam-panel selection}
		% Paragraph 1: Antenna size heterogeneity
		\figref{fig:het_antenna_size} shows mean SE versus beam coherence time for antenna size heterogeneity with $7\times7$ UE antenna panels, where all other antenna parameters match the training configuration. The baseline method is not shown since it only functions with the $3\times3$ training configuration. Both Exhaustive and Hierarchical searches fail to support coherence times below 15ms due to excessive beam training overhead from sweeping 49 beams per panel. In contrast, Static Pred and Traced Path achieve near single-shot beam alignment. Traced Path operates within 0.5 b/s/Hz of the genie-aided method across all coherence times, demonstrating the effectiveness of the path-tracing optimization also shown in \figref{fig:user443_panels}. Static Pred maintains competitive performance, validating that the path predictor successfully maps propagation characteristics agnostic to antenna size variations to the UE locations.
		
		% Paragraph 2: Orientation heterogeneity
		\figref{fig:het_orient} illustrates performance under antenna orientation heterogeneity with $\Delta\alpha = 96^\circ$ while maintaining the $3\times3$ training antenna size. The training configuration uses 9 beams per panel, and we assume the best BS beam is known to isolate UE-side overhead. At the highly mobile scenario with $\Tcoh = 2$ms shown in \figref{fig:het_orient}(b), both Static Pred and Traced Path achieve superior user-for-user SE compared to Baseline. The Pareto frontiers in \figref{fig:het_orient}(a) demonstrate that increasing subset size $\Nbeam$ from 1 to 10 provides diminishing returns, with $\Nbeam=2$ offering an effective balance between overhead and performance. The empirical PDF reveals that Static Pred maintains consistent performance across users despite significant orientation mismatch, confirming that the path information decouples propagation characteristics antenna orientation.

		\begin{figure}
			\centering
			\includegraphics[width=\figsize\linewidth]{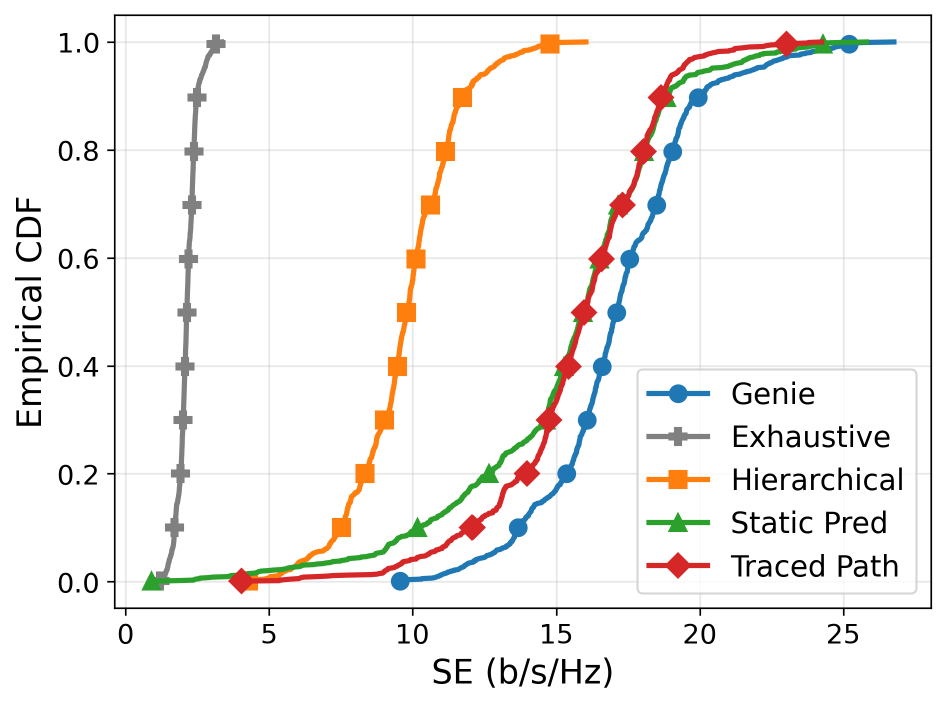}
			\caption{Empirical CDF of SE for antenna size and orientation heterogeneity with $7\times7$ panels, $\Delta\alpha = 96^\circ$, and $\Tcoh = 3$ms. Traced Path is within 1 b/s/Hz of Genie for over 99\% of UEs and Static Pred achieves this for over 70\% of UEs.}
			\label{fig:het_orient_ant_size}
		\end{figure}
		
		% Paragraph 3: Combined antenna size and orientation heterogeneity
		\figref{fig:het_orient_ant_size} presents the empirical CDF of SE under combined antenna size ($7\times7$) and orientation ($\Delta\alpha = 96^\circ$) heterogeneity with $\Tcoh = 3$ms and $\Nbeam=6$. Traced Path achieves performance within 1 b/s/Hz of the genie-aided method for more than 99\% of users, while Static Pred maintains this performance level for more than 70\% of users. The performance gap between Traced Path and Static Pred quantifies the impact of path prediction errors on beam selection quality. Both methods substantially outperform Exhaustive and Hierarchical searches, which cannot support the 3ms coherence time with 49 beams per panel. The results demonstrate that our method generalizes effectively across multiple simultaneous heterogeneity dimensions without retraining.
		
		\begin{figure}
			\centering
			\includegraphics[width=\figsize\linewidth]{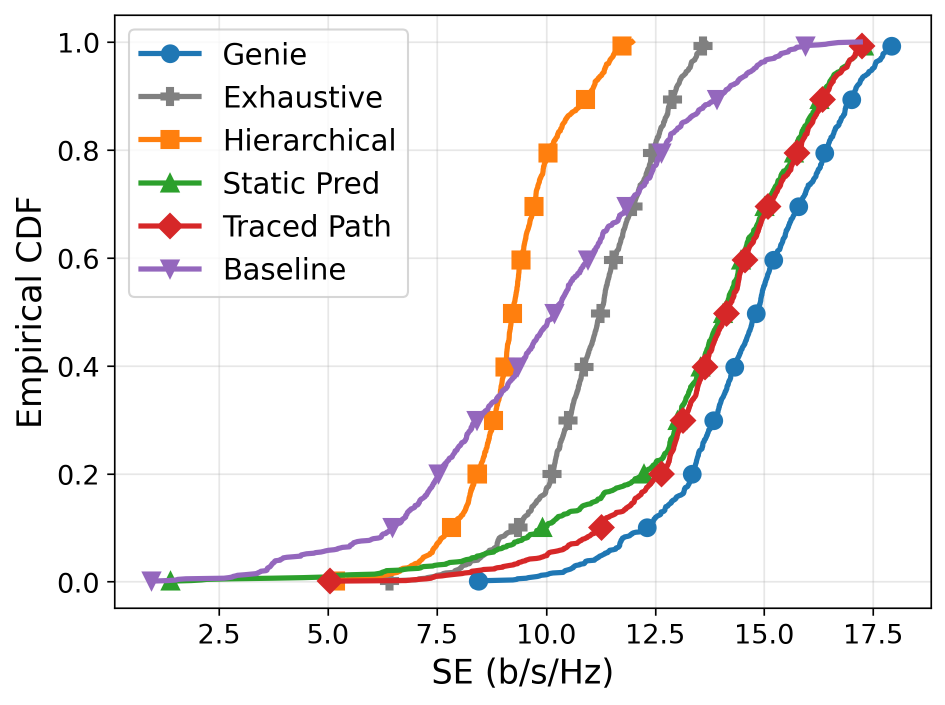}
			\caption{Empirical CDF of SE for orientation and codebook heterogeneity with $3\times3$ panels, $\Delta\alpha = 96^\circ$, $\Tcoh = 2$ms, and DFT test codebooks. Traced Path is within 1 b/s/Hz of Genie for over 99\% of UEs and Static Pred for over 80\%. Baseline performs poorly due to training on angle-limited codebooks.}
			\label{fig:het_orient_codebook}
		\end{figure}

\begin{figure}[ht]
			\centering
			\includegraphics[width=\figsize\linewidth]{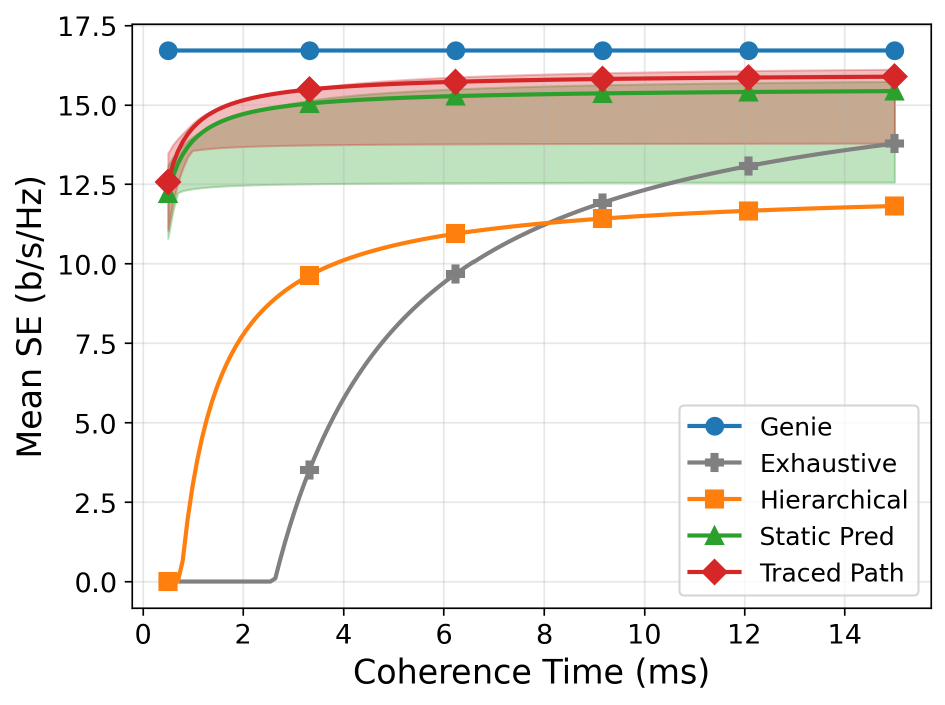}
			\caption{Mean SE vs $\Tcoh$ under antenna size, orientation, and codebook heterogeneity with $7\times7$ panels, $\Delta\alpha = 96^\circ$, $\Tcoh = 3$ms, DFT codebooks, and $\Nbeam = 6$. Our method has a potential to achieve within 0.5 b/s/Hz of the genie-aided method.}
			\label{fig:het_orient_codebook_antenna_size}
		\end{figure}

		% Paragraph 4: Codebook heterogeneity and full heterogeneity
		\figref{fig:het_orient_codebook} and \figref{fig:het_orient_codebook_antenna_size} evaluate performance under codebook heterogeneity, where training used angle-limited codebooks while testing employs DFT codebooks. In \figref{fig:het_orient_codebook}, Traced Path achieves performance within 1 b/s/Hz of the genie for more than 99\% of users, and Static Pred maintains this for more than 80\% of users, while Baseline performs poorly due to codebook-specific training. \figref{fig:het_orient_codebook_antenna_size} presents the most challenging scenario with combined antenna size ($7\times7$), orientation ($\Delta\alpha = 96^\circ$), and codebook heterogeneity. At $\Tcoh = 3$ms with $\Nbeam=6$, Traced Path operates within 0.5 b/s/Hz of the genie-aided method, and Static Pred maintains competitive performance. These results validate that our proposed approach is antenna heterogeneity agnostic and effective across diverse configurations without retraining.

		%			\subsection{Antenna orientation heterogeneity}
		
		%			\subsection{Codebook heterogeneity}
		\section{CONCLUSION}\label{sec:conclusion}
		% \section{Conclusion}\label{sec:conclusion}
		
		In this paper, we addressed the underexplored beam-panel configuration problem for UEs with multi-panel antenna arrays from the perspective of antenna heterogeneity. We developed a heterogeneity-aware RSRP model that decouples propagation characteristics, \textit{path information}, from antenna configuration. This decoupling enables RSRP calculation for arbitrary antenna configurations without retraining. We proposed a location-based path information predictor that maps user locations to propagation characteristics for het-agnostic beam-panel selection.
		
		Simulation results showed that Traced Path achieves near-optimal spectral efficiency within 1 b/s/Hz of genie-aided selection for most users across diverse heterogeneity scenarios including antenna size, orientation, and codebook variations. Static Pred achieves competitive performance with negligible path prediction errors. Our approach promises single-shot beam alignment in highly mobile environments with coherence times as low as 2-3ms, where exhaustive and hierarchical methods fail due to excessive overhead. Our method generalizes effectively across antenna configurations without retraining and is a promising solution for heterogeneity-agnostic beam-panel selection. In future work, we will extend this to multi-user communication, and will investigate polarization effects with channel models providing greater polarization diversity.

		\bibliographystyle{IEEEtran} 
		\bibliography{references_only_paper.bib}
		
		\vfill\pagebreak
		
	\end{document}